\documentclass[11pt,journal,onecolumn]{IEEEtran}
\usepackage{amsfonts}
\usepackage[cmex10]{amsmath}
\usepackage{cite}
\newtheorem{theorem}{Theorem}[section]
\newtheorem{corollary}[theorem]{Corollary}
\newtheorem{lemma}[theorem]{Lemma}
\newtheorem{notation}[theorem]{Notation}

\ifCLASSINFOpdf
\else
\fi
\begin{document}
\bstctlcite{IEEEexample:BSTcontrol}
\title{On a class of left metacyclic codes}

\author{Yonglin Cao,
        Yuan Cao, Fang-Wei Fu
        and~Jian Gao% <-this % stops a space
\thanks{Yonglin Cao is with School of Sciences, Shandong University of Technology, Zibo, Shandong 255091, China. e-mail: ylcao@sdut.edu.cn.}% <-this % stops a space
\thanks{Yuan Cao is with College of Mathematics and Econometrics, Hunan University, Changsha 410082, China. e-mail: yuan$_{-}$cao@hnu.edu.cn.}% <-this % stops a space
\thanks{Fang-Wei Fu is with Chern Institute of Mathematics and LPMC, Nankai University, Tianjin 300071, China. e-mail: fwfu@nankai.edu.cn.}% <-this % stops a space
\thanks{Jian Gao is with School of Sciences, Shandong University of Technology, Zibo, Shandong 255091, China. e-mail: dezhougaojian@163.com.}% <-this % stops a space
\thanks{Manuscript received  , 2016; revised , .}}


\markboth{IEEE Transactions on Information Theory,~Vol.~ , No.~ ,  ~ }%
{Shell \MakeLowercase{\textit{et al.}}: Yonglin Cao, left metacyclic codes}



\maketitle

\begin{abstract}
Let $G_{(m,3,r)}=\langle x,y\mid x^m=1, y^3=1,yx=x^ry\rangle$ be a metacyclic group of order $3m$,
where ${\rm gcd}(m,r)=1$, $1<r<m$ and $r^3\equiv 1$ (mod $m$). Then left ideals of the group algebra $\mathbb{F}_q[G_{(m,3,r)}]$ are called
left metacyclic codes over $\mathbb{F}_q$ of length $3m$, and abbreviated as left $G_{(m,3,r)}$-codes.
A system theory for left $G_{(m,3,r)}$-codes is developed for the case of ${\rm gcd}(m,q)=1$ and $r\equiv q^\epsilon$ for some positive integer $\epsilon$, only using finite field theory and basic theory of cyclic codes and skew
cyclic codes. The fact that any left $G_{(m,3,r)}$-code is a direct sum of concatenated codes with inner codes ${\cal A}_i$ and outer codes $C_i$ is proved, where ${\cal A}_i$ is a minimal cyclic code over $\mathbb{F}_q$ of length $m$ and $C_i$ is a skew cyclic code  of length $3$ over an extension field of $\mathbb{F}_q$. Then an explicit expression for each outer code in any concatenated code is provided. Moreover, the dual code of each left $G_{(m,3,r)}$-code is given and self-orthogonal left $G_{(m,3,r)}$-codes are determined.
\end{abstract}


% \fi
\IEEEpeerreviewmaketitle

\section{Introduction}
\label{intro}
\IEEEPARstart{L}{et} $\mathbb{F}_q$ be a finite field of cardinality $q$ and $G$ a group of order $n$. The group algebra $\mathbb{F}_q[G]$ is a vector space over $\mathbb{F}_q$ with basis $G$. Addition, multiplication with
scalars $c\in \mathbb{F}_q$ and multiplication are defined by: for any $a_g,b_g\in \mathbb{F}_q$,
\begin{center} $\sum_{g\in G}a_g g+\sum_{g\in G}b_g g=\sum_{g\in G}(a_g+b_g) g$, \end{center}

\begin{center} $c(\sum_{g\in G}a_g g)=\sum_{g\in G}ca_g g$, \end{center}

\begin{center} $(\sum_{g\in G}a_g g)(\sum_{g\in G}b_g g)=\sum_{g\in G}(\sum_{uv=g}a_ub_v)g$. \end{center}

\noindent
   Then $\mathbb{F}_q[G]$ is an associative $\mathbb{F}_q$-algebra with identity $1=1_{\mathbb{F}_q}1_{G}$ where
$1_{\mathbb{F}_q}$ and $1_{G}$ are the identity elements of $\mathbb{F}_q$ and $G$ respectively. Readers are referred to \cite{passman2011algebraic}
for more details on group algebra.

\par
  Let $G=G_{(m,s,r)}=\langle x,y\mid x^m=1, y^s=1,yx=x^ry\rangle$ where ${\rm gcd}(m,r)=1$ and $r^s\equiv 1$ (mod $m$).
Then $G$ is called a \textit{metacyclic group} of order $sm$.  Sabin and Lomonaco \cite{sabin1995metacyclic} provided a unique direct decomposition of the $\mathbb{F}_2$-algebra $\mathbb{F}_2[G_{(m,n,r)}]$ to minimal
two-sided ideals (central codes) and described a technique to decompose minimal central codes to
a direct sum of $n$ minimal left ideals (left codes)
and  gave an algorithm to determine these minima left codes. They discovered several good metacyclic codes
and they expressed the hope that more ``good" and perhaps even ``best" codes may be discovered among the ideals of
non-abelian group rings. Recently, Olteanu et al \cite{olteanu2015construction} provided algorithms to construct minimal left
group codes and rediscovered some best codes. These are based on results describing a complete set of orthogonal primitive idempotents in each
Wedderburn component of a semisimple finite group algebra $\mathbb{F}_q[G]$ for a large class of groups $G$. For example, by use of
the computer algebra system GAP and the packages GUAVA and Wedderga some optimal codes and non-abelian group codes were obtained:

\par
  $\diamond$ A linear $[27,18,2]$-code constructed by a left ideal in $\mathbb{F}_2[G]$, where $G=\langle x,y\mid x^9=1, y^3=1,yx=x^4y\rangle$
( \cite[Example 1]{olteanu2015construction}).

\par
  $\diamond$ A best linear $[20,4,8]$-code constructed by a left ideal in $\mathbb{F}_3[G]$, where $G=\langle x,y\mid x^5=1, y^4=1,yx=x^2y\rangle$
(\cite[Example 3]{olteanu2015construction}).

\par
  $\diamond$ A non-abelian group code $[55,10,22]$-code constructed by a left ideal in $\mathbb{F}_2[G]$, where $G=\langle x,y\mid x^{11}=1, y^5=1,yx=x^3y\rangle$
(\cite[Example 4]{olteanu2015construction}).

\par
   For any $\textbf{a}=(a_{0,0}, a_{1,0},\ldots, a_{m-1,0},a_{0,1}, a_{1,1},\ldots, a_{m-1,1}$,
$a_{0,s-1}, a_{1,s-1},\ldots,
a_{m-1,s-1})\in \mathbb{F}_q^{sm}$, define
$$\Psi(\textbf{a})=(1,x,\ldots,x^{m-1})M_{\textbf{a}}\left(\begin{array}{c} 1 \cr y\cr \ldots\cr y^{s-1}\end{array}\right),$$
where
$M_{\textbf{a}}=\left(\begin{array}{cccc} a_{0,0} & a_{0,1}& \ldots &a_{0,s-1} \cr a_{1,0} & a_{1,1} & \ldots & a_{1,s-1}
\cr \ldots & \ldots \cr
a_{m-1,0} & a_{m-1,1}&\ldots & a_{m-1,s-1}\end{array}\right).$
Then $\Psi$ is an $\mathbb{F}_q$-linear isomorphism from $\mathbb{F}_q^{sm}$ onto
$\mathbb{F}_q[G_{(m,s,r)}]$. As in \cite{olteanu2015construction} and \cite{sabin1995metacyclic} a nonempty subset $C$ of $\mathbb{F}_q^{sm}$
is called a \textit{left metacyclic code} (or \textit{left $G_{(m,s,r)}$-code} for more precisely) over $\mathbb{F}_q$ if $\Psi(C)$ is a left ideal of the
$\mathbb{F}_q$-algebra $\mathbb{F}_q[G_{(m,s,r)}]$.
From now on, we will identify $C$ with $\Psi(C)$ for convenience.

\par
   In this paper, we focus our attention on the case of $s=3$ in the
metacyclic group $G_{(m,s,r)}$ and $r=q^\epsilon$ for some positive integer $\epsilon$.

\par
  Compared with the known theory for cyclic codes over finite fields, literatures related with metacyclic codes were
involved too much group algebra language and techniques. A system and elementary theory for left metacyclic codes over finite fields
have not been developed fully to the best of our knowledge. In this paper, we try to achieve the following goals:

\par
   $\bullet$ Develop a system theory for left $G_{(m,3,r)}$-codes using an elementary method. Specifically, only finite field theory and basic theory of cyclic codes and  skew cyclic codes are involved, and it does not involve any group algebra language and technique.

\par
   $\bullet$ Provide a clear expression for each left $G_{(m,3,r)}$-code over $\mathbb{F}_q$ and give a formula to count the number of all such codes.

\par
   $\bullet$ Give an explicit expression of the dual code for each left $G_{(m,3,r)}$-code over $\mathbb{F}_q$ and
determine its self-orthogonality.

\par
   Using the
expression provided, one can list all distinct left $G_{(m,3,r)}$-codes for
specific $m$ and $q$ (not too big) conveniently and easily, design left $G_{(m,3,r)}$-codes for their requirements
and encode the presented codes directly.

\par
   The present paper is organized as follows. In section 2, we prove that any left $G_{(m,3,r)}$-code over $\mathbb{F}_q$
is a direct sum of concatenated codes with inner codes ${\cal A}_i$ and outer codes $C_i$ for the case of ${\rm gcd}(n,q)=1$
and $r\equiv q^\epsilon$ (mod $m$) for some positive integer $\epsilon$, where ${\cal A}_i$ is a minimal cyclic code over $\mathbb{F}_q$ of length $m$ and $C_i$ is a skew $\theta_i$-cyclic code over $K_i$ of length $3$, i.e., left ideals of
the ring $K_i[y;\theta_i]/\langle y^3-1\rangle$, where $K_i$ is an extension field of $\mathbb{F}_q$
and $\theta_i\in {\rm Aut}_{\mathbb{F}_q}(K_i)$ satisfying $\theta_i^3={\rm id}_{K_i}$. In Section 3, we give a precise description for skew $\theta_i$-cyclic codes over $K_i$ of length $3$. Hence all distinct left $G_{(m,3,r)}$-codes over $\mathbb{F}_q$ can be
determined by their concatenated structure. In Section 4, we give the dual code of each left $G_{(m,3,r)}$-code over $\mathbb{F}_q$
precisely and determine all self-orthogonal left-$G_{(m,3,r)}$-codes. Finally, we list all distinct $541696$ left $G_{(14,3,9)}$-codes and
all $3364$ self-orthogonal left $G_{(14,3,9)}$-codes over $\mathbb{F}_3$
in Section 5.

\section{The concatenated structure of left $G_{(m,3,r)}$-codes over $\mathbb{F}_{q}$}

\noindent
In this section, we overview properties for concatenated structure of linear codes and determine the concatenated structure
of left $G_{(m,3,r)}$-codes over $\mathbb{F}_{q}$.

\par
  Let $B$ be a linear $[n_B,k_B,d_B]$-code over $\mathbb{F}_{q}$, $\mathbb{F}_{q^{k_B}}$ an
extension field of $\mathbb{F}_{q}$ with degree $k_B$, $\varphi$ an $\mathbb{F}_{q}$-linear isomorphism from
$\mathbb{F}_{q^{k_B}}$ onto $B$ and $E$ a linear $[n_E,k_E,d_E]$-code over $\mathbb{F}_{q^{k_B}}$.
The \textit{concatenated code} of the inner code $B$ and the outer code $E$ is defined as
$$B\Box_{\varphi}E=\{(\varphi(c_1),\ldots, \varphi(c_{n_E}))\mid (c_1,\ldots,c_{n_E})\in E\}$$
(cf. \cite[Definition 2.1]{sendrier1998concatenated}). It is known that $B\Box_{\varphi}E$ is a linear code over $\mathbb{F}_{q}$ with parameters $[n_Bn_E,k_Bk_E,\geq d_Bd_E]$.

\par
    From now on, let $m$ be a positive integer satisfying ${\rm gcd}(m,q)=1$ and $m\geq 3$, $(\mathbb{Z}_m,+)$ the
addition group of integer residue classes modulo $m$ where $\mathbb{Z}_m=\{0,1,\ldots,m-1\}$, and denote by
$${\cal A}=\mathbb{F}_q[x]/\langle x^m-1\rangle=\{\sum_{i=0}^{m-1}a_ix^i\mid a_0,a_1,\ldots,a_{m-1}\in \mathbb{F}_q\}$$
the residue class ring of $\mathbb{F}_q[x]$ modulo its ideal generated by $x^m-1$ with operations defined
by the usual polynomial operations modulo $x^m-1$. We will identify cyclic codes over
$\mathbb{F}_q$ of length $m$ with ideals of the ring ${\cal A}$ under the identification map:
$(a_0,a_1,\ldots,a_{m-1})\mapsto \sum_{i=0}^{m-1}a_ix^i$.

\par
   First, we  define
$$\theta: (\mathbb{Z}_m,+)\rightarrow (\mathbb{Z}_m,+) \ {\rm by} \ \theta(s)\equiv rs \ ({\rm mod} \ m).$$
As ${\rm gcd}(m,r)=1$, we see that $\theta$ is a group automorphism on $(\mathbb{Z}_m,+)$. Moreover, from
$1<r<n$ and $r^3\equiv 1$ (mod $m$) we deduce that the multiplicative order of $\theta$ is a factor of $3$.

\par
   Next, we define a map ${\cal A}\rightarrow {\cal A}$ by the rule that
$$a(x)\mapsto \sum_{j=0}^{m-1}a_jx^{\theta(j)}\equiv a(x^r) \ ({\rm mod} \ x^m-1),$$
for any $a(x)=\sum_{j=0}^{m-1}a_jx^j\in {\cal A}$. In order to simplify notations, we also use $\theta$ to denote this map on ${\cal A}$, i.e.,
$$\theta(a(x))=a(x^r) \ {\rm in} \ {\cal A}.$$
Then $\theta$ is an $\mathbb{F}_q$-algebra automorphism on ${\cal A}$ satisfying $\theta^3={\rm id}_{{\cal A}}$. In addition, $\theta$
is a permutation of the coordinate positions $\{0,1,\ldots, m-1\}$ of a cyclic codes of length $m$ over $\mathbb{F}_q$ and is called a
\textit{multiplier}. Readers are referred to \cite[Theorems 4.3.12 and 4.3.13]{huffman2010fundamentals}
for more details on basic properties of multipliers.

\par
   Wether $\theta$
denotes this automorphism of ${\cal A}$ or the group automorphism on $(\mathbb{Z}_m,+)$ is determined by the context. Let
$${\cal A}[y;\theta]=\{\sum_{j=0}^ka_j(x)y^j\mid a_0(x),\ldots,a_k(x)\in {\cal A}, \ k\geq 0\}$$
be the \textit{skew polynomial ring} over the commutative ring ${\cal A}$ with coefficients written on the left side, where the multiplication is defined by the rule
$$y^ja(x)=\theta^{j}(a(x))y^j=a(x^{r^j})y^j, \ \forall a(x)\in {\cal A}$$
and by the natural ${\cal A}$-linear extension to all polynomials in ${\cal A}[y;\theta]$.

\par
   As $\theta^3={\rm id}_{\mathcal{A}}$, we have
   $y^3a(x)=a(x)y^3$ for all $a(x)\in {\cal A}$. So $y^3-1$ generates a two-sided ideal $\langle y^3-1\rangle$ of ${\cal A}[y;\theta]$. Let
$${\cal A}[y;\theta]/\langle y^3-1\rangle=\{\alpha+\beta y+\gamma y^2\mid \alpha,\beta,\gamma\in {\cal A}\}$$
be the residue class ring of
$A[y;\theta]$ modulo its two-sided ideal $\langle y^3-1\rangle$. For any $\xi=a_0(x)+a_1(x)y+a_2(x)y^2\in {\cal A}[y;\theta]/\langle y^3-1\rangle$, where
$a_j(x)=\sum_{i=0}^{m-1}a_{i,j}x^i\in {\cal A}$ with $a_{i,j}\in \mathbb{F}_q$ for $j=0,1,2$, we define a natural map:
$$\Phi: \xi\mapsto a(x,y)=\sum_{i=0}^{m-1}\sum_{j=0}^2a_{i,j}x^iy^j.$$
Then it can be easily proved that $\Phi$ is
a ring isomorphism from ${\cal A}[y;\theta]/\langle y^3-1\rangle$ onto $\mathbb{F}_q[G_{(m,3,r)}]$.

\par
   In the rest of this paper,
we will identify $\mathbb{F}_q[G_{(m,3,r)}]$ with ${\cal A}[y;\theta]/\langle y^3-1\rangle$
under this ring isomorphism $\Phi$.

\vskip 3mm
\noindent
 \begin{theorem}\label{th2.1}
 Using the notations above, ${\cal C}$ is a left $G_{(m,3,r)}$-code
over $\mathbb{F}_q$ if and only if ${\cal C}$ is a left ideal of the ring ${\cal A}[y;\theta]/\langle y^3-1\rangle$.
\end{theorem}

\vskip 3mm
\begin{IEEEproof}
    By the identification of
$\mathbb{F}_q[G_{(m,3,r)}]$ with ${\cal A}[y;\theta]/\langle y^3-1\rangle$, we see that ${\cal C}$ is a left $G_{(m,3,r)}$-code
over $\mathbb{F}_q$, i.e., ${\cal C}$ is a left ideal of the ring $\mathbb{F}_q[G_{(m,3,r)}]$,
if and only if ${\cal C}$ is a left ideal of ${\cal A}[y;\theta]/\langle y^3-1\rangle$.
\end{IEEEproof}

\vskip 3mm
\par
  In order to determine all left $G_{(m,3,r)}$-codes
over $\mathbb{F}_q$, by Theorem \ref{th2.1} it is sufficient to give all left ideals of the ring ${\cal A}[y;\theta]/\langle y^3-1\rangle$.
To do this, we need investigate the structure and properties of ${\cal A}$ first.

\par
   For any integer $s$, $0\leq s\leq m-1$, let $J_{s}^{(q)}$ be the $q$-cyclotomic coset modulo $m$, i.e.,
$J_{s}^{(q)}=\{s,sq,\ldots,sq^{l_s-1}\}$ (mod $m$) where $l_s={\rm min}\{k\in \mathbb{Z}^{+}\mid s(q^k-1)\equiv 0 \ ({\rm mod} \ m)\}$.
Then $|J_{s}^{(q)}|=l_s$. It is obvious that $J_{0}^{(q)}=\{0\}$ and  $\theta(J_{0}^{(q)})=J_{0}^{(q)}=\{0\}$.
   In this paper, we assume that
\begin{equation}
  r\equiv q^\epsilon \ ({\rm mod} \ m)
\end{equation}
  for some positive integer $\epsilon$.

\vskip 3mm
\noindent
\begin{lemma}\label{lm2.2}
Let $0\leq s\leq m-1$. Then $J_{s}^{(q)}$ satisfies one and only one of the following
two conditions:

\par
   (I) $\theta(s)\equiv s$ $({\rm mod} \ m)$. In this case, $\theta(k)\equiv k$ $({\rm mod} \ m)$ for all $k\in J_{s}^{(q)}$.

\par
  (II) $\theta(s)\in J_{s}^{(q)}$ and $\theta(s)\not\equiv s$ $({\rm mod} \ m)$. In this case, $l_s=|J_{s}^{(q)}|$ is a multiple of $3$, and $\theta(J_{s}^{(q)})=J_{s}^{(q)}$.
\end{lemma}

\vskip 3mm
\noindent
  \begin{IEEEproof}
  By Condition (1), it follows that $\theta(s)\in J_{s}^{(q)}$ and $\theta(J_{s}^{(q)})=J_{s}^{(q)}$. Then we have one of the following two
cases.

\par
   (I) $\theta(s)=rs\equiv s$ (mod $m$). In this case, for any $k\in J_{s}^{(q)}$ we have
$k\equiv sq^j$ for some $0\leq j\leq l_s-1$, and hence $\theta(k)=rsq^j\equiv sq^j\equiv k$ (mod $m$).

\par
   (II) $\theta(s)\in J_{s}^{(q)}$ and $\theta(s)\not\equiv s$ (mod $m$). In this case, it is obvious
that $l_s=|J_{s}^{(q)}|\geq 2$. By $\theta(s)\in J_{s}^{(q)}$, there exits integer $v$, $1\leq v\leq l_s-1$, such that $\theta(s)=rs\equiv sq^v$ (mod $m$).
  By $r^3\equiv 1$ (mod $m$) and  $sq^v\equiv rs$ (mod $m$), we have
$sq^{3v}=s(q^v)^3\equiv r^3s\equiv s$ (mod $m$). From this and by the minimality
of $l_s$, we deduce that $l_s|3v$.
Suppose that $l_s$ is not a multiple of $3$. Then ${\rm gcd}(l_s,3)=1$ and hence $l_s|v$. By
$sq^{l_s}\equiv s$ (mod $m$), we deduce that $sq^v\equiv s$ (mod $m$), i.e., $\theta(s)\equiv s$ (mod $m$), which contradicts that
$\theta(s)\not\equiv s$ (mod $m$). Hence $l_s$ is a multiple of $3$.
\end{IEEEproof}

\vskip 3mm
\par
    In this paper, let $\zeta$ be a primitive $m$th root of unity in an extension field of $\mathbb{F}_{q}$.
Then $x^m-1=\sum_{s=0}^{m-1}(x-\zeta^s)$. We will adopt the following notations:

\par
  $\bullet$ Let $J_{k_0}^{(q)}, J_{k_1}^{(q)}, \ldots, J_{k_s}^{(q)}$, where $k_0=0$, be all distinct $q$-cyclotomic cosets modulo $m$
satisfying Condition (I) in Lemma \ref{lm2.2}.

\par
   $\bullet$ Let $J_{k_{s+1}}^{(q)}, \ldots, J_{k_{s+t}}^{(q)}$ be all distinct $q$-cyclotomic cosets modulo $m$
satisfying Condition (II) in Lemma \ref{lm2.2}.

\par
  $\bullet$ Denote $J(i)=J_{k_i}^{(q)}$, $f_i(x)=\prod_{j\in J(i)}(x-\zeta^j)$,
  $$K_i=\mathbb{F}_q[x]/\langle f_i(x)\rangle$$
and assume $d_i={\rm deg}(f_i(x))=|J_{k_i}^{(q)}|=|J(i)|$ for all $i=0,1,\ldots,s+t$.

\noindent
  Then $d_i$ is a multiple of $3$ for all $s+1\leq i\leq s+t$
by Lemma \ref{lm2.2}(II).
 It is clear that $f_0(x),f_1(x),\ldots, f_{s+t}(x)$ are
pairwise coprime irreducible polynomials in $\mathbb{F}_q[x]$ satisfying
$$x^m-1=f_0(x)f_1(x)\ldots f_{s+t}(x).$$
Hence $K_i$ is an extension field of
$\mathbb{F}_q$ with cardinality $q^{d_i}$ for $i=0,1,\ldots,s+t$, and $m=\sum_{i=0}^{s+t}d_i$.

\par
  For each integer $i$, $0\leq i\leq s+t$, denote
\begin{center}
$F_i(x)=\frac{x^m-1}{f_i(x)}\in \mathbb{F}_q[x]$.
\end{center}
 Then
${\rm gcd}(F_i(x),f_i(x))$ $=1$. By Extended Euclidian Algorithm, we find polynomials
$u_i(x)$, $v_i(x)\in \mathbb{F}_q[x]$ such that
\begin{equation}
\label{eq2}
 u_i(x)F_i(x)+v_i(x)f_i(x)=1.
\end{equation}
In this paper, we denote
\begin{equation}
\label{eq3}
\varepsilon_i(x)\equiv u_i(x)F_i(x)=1-v_i(x)f_i(x) \ ({\rm mod} \ x^m-1).
\end{equation}
By (\ref{eq2}) and (\ref{eq3}), it follows that $\varepsilon_i(\zeta^j)=1$ for all $j\in J(i)$ and
$\varepsilon_i(\zeta^j)=0$ for all $j\in \mathbb{Z}_m\setminus J(i)$, which implies
\begin{equation}
\label{eq4}
\varepsilon_i(x)=\frac{1}{m}\sum_{l=0}^{m-1}(\sum_{j\in J(i)}\zeta^{-jl})x^l, \ 0\leq i\leq s+t.
\end{equation}  Then we have the following conclusions.

\vskip 3mm
\begin{lemma}\label{lm2.3}
Using the notations above, the following hold.

\par
   (i) $\sum_{i=0}^{s+t}\varepsilon_i(x)=1$, $\varepsilon_i(x)^2=\varepsilon_i(x)$
and $\varepsilon_i(x)\varepsilon_j(x)=0$ in $\mathcal{A}$ for all $0\leq i\neq j\leq s+t$.

\par
   (ii) ${\cal A}={\cal A}_0\oplus {\cal A}_1\oplus\ldots {\cal A}_{s+t}$, where
${\cal A}_i={\cal A}\varepsilon_i(x)$ is the ideal of ${\cal A}$ generated by $\varepsilon_i(x)$,
and ${\cal A}_i$ is a commutative ring with $\varepsilon_i(x)$ as its multiplicative identity for all $i=0,1,\ldots, s+t$.

\par
  (iii) For each integer $i$, $0\leq i\leq s+t$, define
$$\varphi_i: a(x)\mapsto \varepsilon_i(x)a(x) \ {\rm mod} \ x^m-1 \
(\forall a(x)\in K_i).$$
Then $\varphi_i$ is a field isomorphism from $K_i$
onto ${\cal A}_i$.

\par
  (iv) $\theta(\varepsilon_i(x))=\varepsilon_i(x)$ and $\theta({\cal A}_i)={\cal A}_i$, for all $i=0,1$, $\ldots, s+t$.

\par
  (v) For each integer $i$, $0\leq i\leq s+t$, define
$$\theta_i: K_i\rightarrow K_i \
{\rm via} \ a(x)\mapsto a(x^r) \ (\forall a(x)\in K_i).$$
Then $\theta_i$ is an $\mathbb{F}_q$-algebra automorphism on $K_i$
satisfying $\theta_i^3={\rm id}_{K_i}$, where ${\rm id}_{K_i}$ is the identity automorphism on $K_i$. Precisely, $\theta_i={\rm id}_{K_i}$ for all $i=0,1,\ldots,s$, and the multiplicative order of
$\theta_i$ is equal to $3$ for all $i=s+1,\ldots,s+t$.

\par
  (vi) The restriction $\theta|_{{\cal A}_i}$ of $\theta$ on ${\cal A}_i$ is an $\mathbb{F}_q$-algebra automorphism on ${\cal A}_i$
satisfying $(\theta|_{{\cal A}_i})^3={\rm id}_{{\cal A}_i}$ and $\theta |_{{\cal A}_i}=\varphi_i\theta_i\varphi_i^{-1}$. Hence the following diagram for $\mathbb{F}_q$-algebra isomorphisms commutes:
$$\left.\begin{array}{ccc} \ \ \ \  K_i & \stackrel{\theta_i}{\longrightarrow} & K_i
 \cr \varphi_i\downarrow  &    &  \ \ \ \ \downarrow \varphi_i
 \cr  \ \ \ \  \mathcal{A}_i & \stackrel{\theta |_{\mathcal{A}_i}}{\longrightarrow} & \mathcal{A}_i\end{array}\right..$$
Then
$\theta|_{{\cal A}_i}={\rm id}_{{\cal A}_i}$ for all $i=0,1,\ldots,s$, and the multiplicative order of $\theta|_{{\cal A}_i}$ is equal to $3$ for all $i=s+1,\ldots,s+t$.
\end{lemma}

\vskip 3mm
\begin{IEEEproof}
 (i)--(iii) follow from classical ring theory and Equations (\ref{eq2}) and (\ref{eq3}).

\par
  (iv) By the definition of the automorphism $\theta$ on $\mathcal{A}$ and (\ref{eq4}), it follows that
$$\theta(\varepsilon_i(x))=\frac{1}{m}\sum_{l=0}^{m-1}(\sum_{j\in J(i)}\zeta^{-jl})x^{rl \ ({\rm mod} \ m)}.$$
As $J(i)=\theta(J(i))=\{rj \ ({\rm mod} \ m)\mid j\in J(i)\}$, we have
\begin{eqnarray*}
\theta(\varepsilon_i(x))&=&\frac{1}{m}\sum_{l=0}^{m-1}(\sum_{j\in J(i)}\zeta^{-(rj)l})x^{rl \ ({\rm mod} \ m)}\\
 &=&\frac{1}{m}\sum_{l=0}^{m-1}(\sum_{j\in J(i)}\zeta^{-j(rl)})x^{rl \ ({\rm mod} \ m)}.
\end{eqnarray*}
Moreover, since $\theta$ is an automorphism of the group $(\mathbb{Z}_m,+)$, for each $k\in \mathbb{Z}_m$ there is a
unique $l\in \mathbb{Z}_m$ such that $k=rl$ (mod $m$), and hence
$$\theta(\varepsilon_i(x))=\frac{1}{m}\sum_{k=0}^{m-1}(\sum_{j\in J(i)}\zeta^{-jk})x^{k}=\varepsilon_i(x).$$
Therefore,
$\theta({\cal A}_i)=\theta({\cal A})\theta(\varepsilon_i(x))={\cal A}\varepsilon_i(x)
={\cal A}_i.$

\par
  (v) It is clear that $\theta_i$ is an $\mathbb{F}_q$-algebra endomorphism of $K_i$. By $r^3\equiv 1$ (mod $m$) and $f_i(x)=\prod_{j\in J(i)}(x-\zeta^j)$, we see that $(\zeta^j)^{r^3}=\zeta^j$ for all
$j\in J(i)$, which implies that $x^{r^3}\equiv x$ (mod $f_i(x)$), i.e., $x^{r^3}=x$ in $K_i$.
 Hence for any $a(x)\in K_i=\mathbb{F}_q[x]/\langle f_i(x)\rangle$
we have $\theta_i^3(a(x))=a(x^{r^3})=a(x)$ in $K_i$. Therefore, $\theta_i^3={\rm id}_{K_i}$ and
so $\theta_i$ is an $\mathbb{F}_q$-algebra automorphism of $K_i$.
Then we consider the following two cases:

\par
    (v-1) Let $0\leq i\leq s$. For any $j\in J(i)$, by Condition (I) in Lemma 2.2 we have
$(\zeta^j)^r=\zeta^j$, which implies  $x^r\equiv x$ (mod $x-\zeta^j$), and hence
$x^r\equiv x$ (mod $f_i(x)$). Therefore, $\theta_i(a(x))=a(x^r)=a(x)$ for any $a(x)\in K_i=\mathbb{F}_q[x]/\langle f_i(x)\rangle$, and so $\theta_i={\rm id}_{K_i}$.

\par
   (v-2) Let $s+1\leq i\leq s+t$. By Condition (II) in Lemma 2.2, there exists
$j\in J(i)$ such that $rj\not\equiv j$ (mod $m$), which implies $(\zeta^j)^r\neq\zeta^j$.
Then by the proof of (v-1), we conclude $\theta_i\neq{\rm id}_{K_i}$. From this and by $\theta_i^3={\rm id}_{K_i}$,
we deduce that the multiplicative order of
$\theta_i$ is equal to $3$.

\par
  (vi) By (iv) and $\theta^3={\rm id}_{{\cal A}}$, it follows that $\theta|_{{\cal A}_i}$ is an $\mathbb{F}_q$-algebra automorphism on ${\cal A}_i$
satisfying $(\theta|_{{\cal A}_i})^3={\rm id}_{{\cal A}_i}$.
Then the equation $\theta |_{{\cal A}_i}=\varphi_i\theta_i\varphi_i^{-1}$ follows from (v) and the definitions of
$\varphi_i$, $\theta_i$ and $\theta|_{{\cal A}_i}$ immediately.
\end{IEEEproof}

\vskip 3mm
\par
   For any integer $i$, $0\leq i\leq s+t$, it is known that
$\mathcal{A}_i$ is a minimal cyclic code of length $m$ over $\mathbb{F}_q$.
Precisely, $f_i(x)$ is the parity check polynomial and $\varepsilon_i(x)$ is the idempotent generator of $\mathcal{A}_i$.
Hence ${\rm dim}_{\mathbb{F}_q}(\mathcal{A}_i)={\rm deg}(f_i(x))=d_i$.

\par
   Frow now on, we adopt the following notations.

\par
  $\bullet$ Let $K_i[y;\theta_i]=\{\sum_{j=0}^kb_j(x)y^j\mid b_0(x),\ldots,b_k(x)\in K_i, \ k\geq 0\}$
be the skew polynomial ring over $K_i$ with coefficients written on the left side, where the multiplication is defined by the rule
$$y^ja(x)=\theta_i^{j}(a(x))y^j=a(x^{r^j})y^j, \ \forall a(x)\in K_i$$
and by the natural $K_i$-linear extension to all polynomials in $K_i[y;\theta_i]$.

\par
   Since $\theta_i^3={\rm id}_{K_i}$ by Lemma \ref{lm2.3}(v), we see that $y^3-1$ generates a two-sided ideal $\langle y^3-1\rangle$
of $K_i[y;\theta_i]$.

\par
  $\bullet$ Let $K_i[y;\theta_i]/\langle y^3-1\rangle=\{a(x)+b(x)y+c(x)y^2\mid a(x), b(x),c(x)\in K_i\}$
be the residue class ring of
$K_i[y;\theta_i]$ modulo its two-sided ideal $\langle y^3-1\rangle$.

\par
   Recall that left ideals
of $K_i[y;\theta_i]/\langle y^3-1\rangle$ are called \textit{skew $\theta_i$-cyclic codes}
over $K_i$ of length $3$ (See \cite[Theorem 1]{boucher2007skew}).
For more details on skew cyclic codes, readers are referred to \cite{boucher2007skew, boucher2008skew, boucher2009codes, boucher2009coding, Jitman201239}.

\par
   Now, we can decompose any left $G_{(m,3,r)}$-code into a direct sum of concatenated codes
by the following theorem.

\vskip 3mm
\begin{theorem}\label{th2.4}
Using the notations above, we have the following conclusions.

\par
  (i) ${\cal A}[y;\theta]/\langle y^3-1\rangle=\oplus_{i=0}^{s+t}({\cal A}_i[y;\theta |_{{\cal A}_i}]/\langle \varepsilon_i(x)y^3-\varepsilon_i(x)\rangle)$.

\par
  (ii) For each integer $i$, $0\leq i\leq s+t$, $\varphi_i:K_i\rightarrow {\cal A}_i$ can be extended to a ring isomorphism from
$K_i[y;\theta_i]/\langle y^3-1\rangle$ onto ${\cal A}_i[y;\theta |_{{\cal A}_i}]/\langle \varepsilon_i(x)y^3-\varepsilon_i(x)\rangle$ by
$$\varphi_i:\xi(y)\mapsto \varphi_i(\xi_0)+\varphi_i(\xi_1)y+\varphi_i(\xi_2)y^2=\varepsilon_i(x)\xi(y)$$
for any $\xi(y)=\xi_0+\xi_1+\xi_2y^2$ with $\xi_0,\xi_1,\xi_2\in K_i$.

\par
  (iii) ${\cal C}$ is a left $G_{(m,3,r)}$-code over $\mathbb{F}_q$ if and only if for each
$0\leq i\leq s+t$ there is a unique skew $\theta_i$-cyclic code $C_i$ over $K_i$
of length $3$ such that
$${\cal C}=\bigoplus_{i=0}^{s+t}{\cal A}_i\Box_{\varphi_i}C_i,$$
where
${\cal A}_i\Box_{\varphi_i}C_i=\{\varepsilon_i(x)\xi \ ({\rm mod} \ x^m-1)\mid \xi\in C_i\}$. Moreover, we have $|{\cal C}|=\prod_{i=0}^{s+t}|C_i|$.
\end{theorem}

\vskip 3mm
\begin{IEEEproof} (i) By Lemma \ref{lm2.3}(ii),(iv) and (vi), we have
 $${\cal A}[y;\theta]=\oplus_{i=0}^{s+t}{\cal A}_i[y;\theta |_{{\cal A}_i}].$$
Moreover,  by ${\cal A}_i={\cal A}\varepsilon_i(x)$ we know that the projection from ${\cal A}[y;\theta]$ onto ${\cal A}_i[y;\theta |_{{\cal A}_i}]$ is determined by $\alpha(y)\mapsto \varepsilon_i(x)\alpha(y)$ ($\forall \alpha(y)\in {\cal A}[y;\theta]$). Especially, we have $y^3-1\mapsto \varepsilon_i(x)y^3-\varepsilon_i(x)$ under this projection.
   As $(\theta |_{{\cal A}_i})^3={\rm id}_{{\cal A}_i}$, $\varepsilon_i(x)y^3-\varepsilon_i(x)$ generates a two-sided ideal
$\langle \varepsilon_i(x)y^3-\varepsilon_i(x)\rangle$ of ${\cal A}_i[y;\theta |_{{\cal A}_i}]$
for all $i=0,1,\ldots,s+t$.
By Lemma \ref{lm2.3}(i) it follows that $y^3-1=\sum_{i=0}^{s+t}(\varepsilon_i(x)y^3-\varepsilon_i(x))$ and $(\varepsilon_i(x)y^3-\varepsilon_i(x))(\varepsilon_j(x)y^3-\varepsilon_j(x))=0$ for all $0\leq i\neq j\leq s+t$.
Hence
\begin{center}
${\cal A}[y;\theta]/\langle y^3-1\rangle=\oplus_{i=0}^{s+t}({\cal A}_i[y;\theta |_{{\cal A}_i}]/\langle \varepsilon_i(x)y^3-\varepsilon_i(x)\rangle)$.
\end{center}

\par
   (ii) Since $\varphi_i:K_i\rightarrow {\cal A}_i$ is a ring isomorphism by Lemma 2.3(iii), the conclusion follows
 from Lemma \ref{lm2.3}(vi) and a direct calculation.

\par
   (iii) By Theorem \ref{th2.1} and (i), we see that ${\cal C}$ is a left $G_{(m,3,r)}$-code over $\mathbb{F}_q$ if and only if
for each integer $i$,
$0\leq i\leq s+t$, there is a unique left ideal ${\cal C}_i$ of the ring ${\cal A}_i[y;\theta |_{{\cal A}_i}]/\langle \varepsilon_i(x)y^3-\varepsilon_i(x)\rangle$
such that ${\cal C}=\oplus_{i=0}^{s+t}{\cal C}_i$.
By (ii), the latter condition is equivalent to that for each integer $i$, $0\leq i\leq s+t$, there is a unique left ideal $C_i$ of $K_i[y;\theta_i]/\langle y^3-1\rangle$ such that
${\cal C}_i=\varphi_i(C_i)=\{\varepsilon_i(x)\xi\mid \xi\in C_i\}={\cal A}_i\Box_{\varphi_i}C_i$.

\par
 Finally, it is clear that the codewords contained in ${\cal C}$ is equal to $|{\cal C}|=\prod_{i=0}^{s+t}|{\cal C}_i|=\prod_{i=0}^{s+t}|C_i|$.
\end{IEEEproof}

\vskip 3mm
\par
  By Theorem \ref{th2.4}, in order to give all distinct left $G_{(m,3,r)}$-codes over $\mathbb{F}_q$ it is sufficient to
determine all distinct skew $\theta_i$-cyclic codes
over $K_i$ of length $3$, for all $i=0,1,\ldots,s+t$. For convenience and notations simplicity in the following sections, we introduce
the following notations.

\vskip 3mm
\noindent
\begin{notation} For any integer $i$, $0\leq i\leq s+t$, denote

\par
  $\bullet$ ${\cal R}_i=K_i[y;\theta_i]/\langle y^3-1\rangle$.

\noindent
  For any $\alpha=a_0+a_1y+a_2y^2\in {\cal R}_i$ with $a_0,a_1,a_2\in K_i$, the \textit{Hamming weight}
${\rm wt}_H^{(i)}(\alpha)$ of $\alpha$ over $K_i$ is defined as the number of nonzero coefficients of
the polynomial $a_0+a_1y+a_2y^2\in K_i[y;\theta_i]$, i.e.,

\par
  $\bullet$ ${\rm wt}_H^{(i)}(\alpha)=|\{j\mid a_j\neq0, \ j=0,1,2\}|$.

\noindent
   For any nonzero left ideal $J$ of ${\cal R}_i$,
the \textit{minimum Hamming weight}
${\rm wt}_H^{(i)}(J)$ of $J$ over $K_i$ is defined as

\par
  $\bullet$  ${\rm wt}_H^{(i)}(J)={\rm min}\{{\rm wt}_H^{(i)}(\alpha)\mid \alpha\neq 0, \ \alpha\in J\}$.
\end{notation}

%%%%%%%%%%%%%%%%%%%%%%%%%%%%%%%%%%%%%%%%%%%%%%%%%%%%%%%%%%%%%%%%%%%%%%%%%%%%%%%%%%%%%%%%%%%%%%%%%%%%%%%

%%%%%%%%%%%%%%%%%%%%%%%%%%%%%%%%%%%%%%%%%%%%%%%%%%%%%%%%%%%%%%%%%%%%%%%%%%%%%%%%%%%%%%%%%%%%%%%%%%%%%%%
\section{Skew $\theta_i$-cyclic codes
over $K_i$ of length $3$}

\noindent
In this section, we give all skew $\theta_i$-cyclic codes
over $K_i$ of length $3$, i.e., left ideals of the ring ${\cal R}_i$,
where $0\leq i\leq s+t$.

\par
  $\diamondsuit$ Let $0\leq i\leq s$. Then $K_i$ is a finite field of cardinality $q^{d_i}$
and $\theta_i$ is the identity automorphism of $K_i$ by Lemma \ref{lm2.3}(v). Hence ${\cal R}_i=K_i[y]/\langle y^3-1\rangle$ which is a commutative ring.
In this case, left ideals
of ${\cal R}_i$ are in fact ideals of ${\cal R}_i$. By the basic theory of cyclic codes over finite fields,
we know that $C_i$ is an ideal of ${\cal R}_i$ if and only if $C_i$ is a cyclic code over $K_i$ of length $3$.
The latter is equivalent to that there is a unique monic divisor $g(y)$ of $y^3-1$ in $K_i[y]$ such
that $C_i={\cal R}_ig(y)$. Then $g(y)$ is called the \textit{generator polynomial} of $C_i$ and
${\rm dim}_{K_i}(C_i)=3-{\rm deg}_y(g(y))$ where ${\rm deg}_y(g(y))$ is
the degree of $g(y)$ as a polynomial with indeterminate $y$. Obviously, ${\cal R}_i(y^3-1)=\{0\}$.

\vskip 3mm
\noindent
\begin{theorem}\label{th3.1}
 Let $0\leq i\leq s$. Then ${\cal R}_i=K_i[y]/\langle y^3-1\rangle$ which is a commutative ring, and the following hold.

\par
  (i) If $q\equiv 0$ $({\rm mod} \ 3)$, there are $4$ distinct ideals in ${\cal R}_i$:
$${\cal R}_ig(y), \ {\rm where} \ g(y)\in \{1, y-1,(y-1)^2,y^3-1\}.$$

  (ii) If $q^{d_i}\equiv 2$ $({\rm mod} \ 3)$, there are $4$ distinct ideals in ${\cal R}_i$:
$${\cal R}_ig(y), \ {\rm where} \ g(y)\in \{1, y-1,y^2+y+1,y^3-1\}.$$

  (iii) Let $q^{d_i}\equiv 1$ $({\rm mod} \ 3)$, $\zeta_i(x)$ a primitive element of $K_i$ and
denote $\omega_i(x)=\zeta_i(x)^{\frac{q^{d_i}-1}{3}}$ $({\rm mod} \ f_i(x))$. Then
there are $8$ distinct ideals
in ${\cal R}_i$: ${\cal R}_ig(y)$, \textit{where}
\begin{eqnarray*}
&&g(y)\in \{1, y-1,y-\omega_i(x),y-\omega_i(x)^2,y^2+y+1,\\
 && \ \ \ \ \  (y-1)(y-\omega_i(x)^2), (y-1)(y-\omega_i(x)), y^3-1\}.
\end{eqnarray*}
\end{theorem}

\vskip 3mm
\begin{IEEEproof}
 (i) Since $q\equiv 0$ (mod $3$), we have $y^3-1=(y-1)^3$ in $K_i[y]$.
In this case, $y^3-1$ has $4$ monic divisors in $K_i[y]$: $1$, $y-1$, $(y-1)^2$ and $y^3-1$,.

\par
  (ii) Since $q^{d_i}\equiv 2$ (mod $3$), we have that $y^3-1=(y-1)(y^2+y+1)$ where
$y^2+y+1$ is irreducible in $K_i[y]$. In this case, $y^3-1$ has $4$ monic divisors in $K_i[y]$: $1$, $y-1$, $y^2+y+1$ and $y^3-1$.

\par
  (iii) Since $q^{d_i}\equiv 1$ (mod $3$), we have $3|(q^{d_i}-1)$ and $\omega_i(x)$ is a primitive $3$th root of unity in $K_i$, which implies
$y^3-1=(y-1)(y-\omega_i(x))(y-\omega_i(x)^2)$. In this case, $y^3-1$ has $8$ monic divisors in $K_i[y]$:  $1$, $y-1$,
$y-\omega_i(x)$, $y-\omega_i(x)^2$, $y^2+y+1$, $(y-1)(y-\omega_i(x)^2)$, $(y-1)(y-\omega_i(x))$ and $y^3-1$.
\end{IEEEproof}

\vskip 3mm
\par
  $\diamondsuit$ Let $s+1\leq i\leq s+t$. Then $K_i$ is a finite field of cardinality $q^{d_i}$, $d_i$ is a multiple of $3$ by Lemma \ref{lm2.2}(II),
and $\theta_i$ is an $\mathbb{F}_q$-algebra automorphism of $K_i$ with multiplicative order $3$ by Lemma \ref{lm2.3}(v).

\par
   In this case, by \cite[Lemma 1 and Theorem 1]{boucher2007skew} we know that $C_i$ is a left ideal of ${\cal R}_i$ if and only if
$C_i$ is a skew $\theta_i$-cyclic code over $K_i$ of length $3$, and the latter is equivalent to that
there is a unique monic right divisor $g(y)$ of $y^3-1$ in the skew polynomial ring $K_i[y;\theta_i]$
such that
$C_i={\cal R}_ig(y)$. If the latter condition is satisfied, $g(y)$ is called the \textit{generator polynomial} of $C_i$,
${\rm dim}_{K_i}(C_i)=3-{\rm deg}_y(g(y))$ and the number
of codewords in $C_i$ is equal to $|C_i|=(q^{d_i})^{3-{\rm deg}_y(g(y))}$. Precisely, a generator matrix of $C_i$ over $K_i$
is given by:
$G_{C_i}=(a(x),b(x),1) \ {\rm if} \
g(x)=a(x)+b(x)y+y^2;$ and
$$G_{C_i}=\left(\begin{array}{ccc}a(x) & 1 & 0 \cr 0 &\theta_i(a(x)) & 1
\end{array}\right) \ {\rm if} \
g(x)=a(x)+y.$$

\par
   In order to describe monic right divisors of $y^3-1$ in the skew polynomial ring $K_i[y;\theta_i]$ and
the relationships between two nontrivial monic right divisors, we adopt the following notations
in the rest of this paper:

\par
  $\bullet$ Let $\zeta_i(x)$ be a primitive element of $K_i$
and denote $\varrho_i(x)=\zeta_i(x)^{q^{\frac{d_i}{3}-1}}$. Then the multiplicative order of $\varrho_i(x)$ is equal to
${\rm ord}(\varrho_i(x))=\frac{q^{d_i}-1}{q^{\frac{d_i}{3}}-1}=1+q^{\frac{d_i}{3}}+q^{\frac{2d_i}{3}}.$
Denote
$$\mathcal{G}_i=\{\varrho_i(x)^k\mid k=0,1,\ldots,q^{\frac{d_i}{3}}+q^{\frac{2d_i}{3}}\}$$
which is the multiplicative cyclic subgroup
of $K_i^\times$ generated by $\varrho_i(x)$. Then $|\mathcal{G}_i|=1+q^{\frac{d_i}{3}}+q^{\frac{2d_i}{3}}$.

\vskip 3mm
\begin{lemma}\label{lm3.2}
Let $s+1\leq i\leq s+t$. Then we have the following:

\par
   (i) $\theta_i(a(x))=a(x)^{q^{\frac{d_i}{3}}}$ for all $a(x)\in K_i$ or $\theta_i(a(x))=a(x)^{q^{\frac{2d_i}{3}}}$ for all $a(x)\in K_i$.

\par
  (ii) For any $\alpha(x)\in K_i^\times$, $\alpha(x)\in \mathcal{G}_i$ if and only if $\alpha(x)$ satisfies the following equation
$$\alpha(x)\theta_i(\alpha(x))\theta_i^2(\alpha(x))=1.$$

\par
   (iii) All distinct monic right divisors of $y^3-1$ with degree $1$ in the skew polynomial ring $K_i[y;\theta_i]$
are given by:
$$-\alpha(x)+y, \ \alpha(x)\in \mathcal{G}_i.$$

\par
   (iv) All distinct monic right divisors of $y^3-1$ with degree $2$ in the skew polynomial ring $K_i[y;\theta_i]$
are given by:
$$\alpha(x)^{-1}+\theta_i^2(\alpha(x))y+y^2, \
\alpha(x)\in \mathcal{G}_i.$$

\par
   (v) For any $\alpha(x)\in \mathcal{G}_i$, we have
\begin{eqnarray*}
y^3-1&=&\left(-\alpha(x)+y\right)\cdot\left(\alpha(x)^{-1}+\theta_i^2(\alpha(x))y+y^2\right)\\
&=&\left(\alpha(x)^{-1}+\theta_i^2(\alpha(x))y+y^2\right)\cdot\left(-\alpha(x)+y\right).
\end{eqnarray*}

\noindent
Therefore, both the number of right divisors of $y^3-1$ in $K_i[y;\theta_i]$ with degree $1$
and the number of right divisors of $y^3-1$ in $K_i[y;\theta_i]$ with degree $2$ are equal to $1+q^{\frac{d_i}{3}}+q^{\frac{2d_i}{3}}$.
\end{lemma}

\vskip 3mm
\begin{IEEEproof}
(i) Let $\sigma: \alpha\mapsto \alpha^q$ ($\forall \alpha\in K_i$) be the Frobenius automorphism of $K_i$ over $\mathbb{F}_q$.
Then the multiplicative order of $\sigma$ is $d_i$ and every automorphism of $K_i$ over $\mathbb{F}_q$ is of the form: $\sigma^k$, $0\leq k\leq d_i-1$. By Lemma \ref{lm2.2}, $d_i$ is a multiple of $3$. Hence ${\rm ord}(\sigma^k)=3$ if and only if
${\rm gcd}(k,d_i)=\frac{d_i}{3}$, i.e., $k=\frac{d_i}{3}$ or $k=\frac{2d_i}{3}$.
By Lemma \ref{lm2.3}(v), $\theta_i$ is an automorphism of $K_i$ over $\mathbb{F}_q$ with multiplicative order $3$,
which implies that $\theta_i=\sigma^{\frac{d_i}{3}}$ or $\theta_i=\sigma^{\frac{2d_i}{3}}$. Hence
$\theta_i(a(x))=a(x)^{q^{\frac{d_i}{3}}}$ for all $a(x)\in K_i$ or $\theta_i(a(x))=a(x)^{q^{\frac{2d_i}{3}}}$ for all $a(x)\in K_i$.

\par
  (ii) Denote $\alpha=\alpha(x)$ in order to simplify the notation. When $\theta_i(a(x))=a(x)^{q^{\frac{d_i}{3}}}$ for all $a(x)\in K_i$,
it is clear that
$$\alpha\theta_i(\alpha)\theta_i^2(\alpha)=\alpha^{1+q^{\frac{d_i}{3}}+q^{\frac{2d_i}{3}}}.$$
When $\theta_i(a(x))=a(x)^{q^{\frac{2d_i}{3}}}$ for all $a(x)\in K_i$, by $\alpha^{q^{\frac{4d_i}{3}}}=(\alpha^{q^{d_i}})^{q^{\frac{d_i}{3}}}
=\alpha^{q^{\frac{d_i}{3}}}$ we have
$\alpha\theta_i(\alpha)\theta_i^2(\alpha)=\alpha^{1+q^{\frac{2d_i}{3}}+q^{\frac{4d_i}{3}}}=\alpha^{1+q^{\frac{d_i}{3}}+q^{\frac{2d_i}{3}}}$ as well.

\par
   Since $K_i^\times$ is a multiplicative cyclic group with order $q^{d_i}-1$ and $\mathcal{G}_i$
is a subgroup of $K_i^\times$ with order $1+q^{\frac{d_i}{3}}+q^{\frac{2d_i}{3}}$, by basic group theory
we conclude that $\alpha\in \mathcal{G}_i$ if and only if $\alpha^{1+q^{\frac{d_i}{3}}+q^{\frac{2d_i}{3}}}=1$, i.e.,
$\alpha\theta_i(\alpha)\theta_i^2(\alpha)=1$.

\par
  (iii) Let $\alpha\in K_i$. Dividing $y^3-1$ by $y-\alpha$ from right hand in the skew polynomial ring $K_i[y;\theta_i]$, we
\begin{eqnarray*}
y^3-1&=&\left(y^2+\theta_i^2(\alpha)y+\theta_i^2(\alpha)\theta_i(\alpha)\right)(y-\alpha)\\
  &&+1-\alpha\theta_i(\alpha)\theta_i^2(\alpha).
\end{eqnarray*}
Hence $-\alpha+y$ is a right divisors of $y^3-1$ in $K_i[y;\theta_i]$ if and only if
$1-\alpha\theta_i(\alpha)\theta_i^2(\alpha)=0$, i.e., $\alpha\theta_i(\alpha)\theta_i^2(\alpha)=1$.

\par
  (iv) Let $\beta,\gamma\in K_i$. Then $\gamma+\beta y+y^2$ is a right divisors of $y^3-1$ in $K_i[y;\theta_i]$
if and only if there exists $\alpha\in K_i$ such that
\begin{eqnarray*}
y^3-1&=&(-\alpha+y)(\gamma+\beta y+y^2)\\
  &=&y^3+(\theta_i(\beta)-\alpha)y^2+(\theta_i(\gamma)-\beta\alpha)y-\gamma \alpha,
\end{eqnarray*}
which is equivalent that $\alpha,\beta,\gamma$ satisfy
$\theta_i(\beta)=\alpha$, $\theta_i(\gamma)=\beta\alpha$  and $\gamma \alpha=1$. From these and by $\theta_i^3={\rm id}_{K_i}$, we deduce
that
$\beta=\theta_i^2(\theta_i(\beta))=\theta_i^2(\alpha)$ and $\gamma=\theta_i^2(\theta_i(\gamma))=\theta_i^2(\beta\alpha)=\theta_i(\alpha)\theta_i^2(\alpha)$,
which implies $\alpha\theta_i(\alpha)\theta_i^2(\alpha)=1$.
Then by the latter equation and (ii), we conclude that
$\alpha\in\mathcal{G}_i$,
and hence
$$\gamma=\alpha^{-1} \ {\rm and} \ \beta=\theta_i^2(\alpha).$$

\par
  Therefore, both the number of right divisors of $y^3-1$ in $K_i[y;\theta_i]$ with degree $1$
and the number of right divisors of $y^3-1$ in $K_i[y;\theta_i]$ with degree $2$ are equal to $1+q^{\frac{d_i}{3}}+q^{\frac{2d_i}{3}}$.

\par
  (v) It follow from a direct calculation.
\end{IEEEproof}

\vskip 3mm
\par
   Now, by Lemma \ref{lm3.2} and basic theory of skew cyclic codes we list all distinct left ideals of ${\cal R}_i$, i.e., skew $\theta_i$-cyclic codes
over $K_i$ of length $3$, by the following theorem.

\vskip 3mm
\begin{theorem}\label{th3.3}
Let $s+1\leq i\leq s+t$. Then all distinct left ideals of ${\cal R}_i=K_i[y;\theta_i]/\langle y^3-1\rangle$ are given by one of the following
three cases:

\par
  (i) $C_{i,0}=\{0\}$; $C_{i,3}={\cal R}_i$ with $|C_{i,3}|=q^{3d_i}$, ${\rm wt}_H^{(i)}(C_{i,3})=1$ and $G_{C_{i,3}}=I_3$
is a generator matrix of $C_{i,3}$, where $I_3$ is the identity matrix of size $3\times 3$.

\par
  (ii) $C_{i,2,\alpha}={\cal R}_i(-\alpha(x)+y)$, where  $\alpha=\alpha(x)\in \mathcal{G}_i$.

\par
  Precisely, we have ${\rm dim}_{K_i}(C_{i,2,\alpha})=2$, $|C_{i,2,\alpha}|=q^{2d_i}$ and a generator matrix of $C_{i,2,\alpha}$ is given by
$$G_{C_{i,2,\alpha}}=\left(\begin{array}{ccc}-\alpha(x) & 1 & 0\cr
0 & -\theta_i(\alpha(x))& 1\end{array}\right).$$
Therefore, $C_{i,2,\alpha}=\{(a,b)G_{C_{i,2,\alpha}}\mid a,b\in K_i\}$ and
${\rm wt}_H^{(i)}(C_{i,2,\alpha})=2$. Hence $C_{i,2,\alpha}$ is a MDS linear $[3,2,2]_{q^{d_i}}$-code over $K_i$.

\par
  (iii) $C_{i,1,\alpha}={\cal R}_i(\alpha(x)^{-1}+\theta_i^2(\alpha(x))y+y^2)$, where
$\alpha=\alpha(x)\in\mathcal{G}_i$.

\par
  Precisely, we have ${\rm dim}_{K_i}(C_{i,1,\alpha})=1$, $|C_{i,1,\alpha}|=q^{d_i}$ and a generator matrix of $C_{i,1,\alpha}$ is given by
$$G_{C_{i,1,\alpha}}=(\alpha(x)^{-1},\theta_i^2(\alpha(x)),1).$$
Hence $C_{i,1,\alpha}=\{a G_{C_{i,1,\alpha}}\mid a \in K_i\}$ and
${\rm wt}_H^{(i)}(C_{i,1,\alpha})=3$. Then $C_{i,1,\alpha}$ is a MDS linear $[3,1,3]_{q^{d_i}}$-code over $K_i$.

\par
  Therefore, the number of left ideals
of ${\cal R}_i$ is equal to $4+2q^{\frac{d_i}{3}}+2q^{\frac{2d_i}{3}}$.
\end{theorem}

\vskip 3mm
\par
  In the rest of this paper, for any $\alpha,\beta\in \mathcal{G}_i$, we denote

\vskip 2mm\par
  $\bullet$ $C_{i,2,\alpha}=\mathcal{R}_i(-\alpha+y)$,
$C_{i,1,\beta}={\cal R}_i(\beta^{-1}+\theta_i^2(\beta)y+y^2)$

\vskip 2mm\noindent
which are skew $\theta_i$-cyclic codes over $K_i$ of length $3$.
   By Theorems \ref{th2.4}, \ref{th3.1} and \ref{th3.3}, we deduce the following corollary.

\vskip 3mm
\begin{corollary}\label{co3.4}
Using the notations in Section 2, denote
$$\delta=|\{d_i \mid q^{d_i}\equiv 1 \ ({\rm mod} \ 3), \ 0\leq i\leq s\}|.$$
Then the number $N_{(m,3,r;q)}$ of left $G_{(m,3,r)}$-codes over $\mathbb{F}_q$ is given by
one of the following two cases:

\par
  (i) When $q\equiv 0$ $($mod $3)$,
$$N_{(m,3,r;q)}=4^{s+1}\prod_{i=s+1}^{s+t}(4+2q^{\frac{d_i}{3}}+2q^{\frac{2d_i}{3}}).$$

\par
  (ii) When $q\not\equiv 0$ $($mod $3)$,
$$N_{(m,3,r;q)}=2^{\delta}4^{s+1}\prod_{i=s+1}^{s+t}(4+2q^{\frac{d_i}{3}}+2q^{\frac{2d_i}{3}}).$$
\end{corollary}

\vskip 3mm\par
  As the end of this section, we investigate the relationship between two left ideals of the ring $\mathcal{R}_i$ for
$i=s+1,\ldots,s+t$. To do this, we consider the relationship between two nontrivial monic right divisors
of $y^3-1$ in $K_i[y,\theta_i]$ with different degrees.

\vskip 3mm
\begin{lemma}\label{lm3.5}
For any $\gamma\in \mathcal{G}_i$, denote
$$\phi_{i,\gamma}(X)=X^{q^{\frac{d_i}{3}}+1}+\gamma^{q^{\frac{2d_i}{3}}}X+\gamma^{-1},$$
$$\psi_{i,\gamma}(X)=\gamma^{q^{\frac{d_i}{3}}}X^{q^{\frac{d_i}{3}}+1}+\gamma^{-1}X^{q^{\frac{d_i}{3}}}+1.$$
Then both $\phi_{i,\gamma}(X)$ and $\psi_{i,\gamma}(X)$ have exactly $q^{\frac{d_i}{3}}+1$
roots and these roots are contained in $\mathcal{G}_i$.
\end{lemma}

\vskip 3mm
  \begin{IEEEproof}
  As ${\rm gcd}(q,q^{\frac{d_i}{3}}+1)=1$,
$\phi_{i,\gamma}(X)$ and $\psi_{i,\gamma}(X)$ have exactly $q^{\frac{d_i}{3}}+1$
roots in some extended field of $K_i$. We only need to prove that
all these roots are contained in $\mathcal{G}_i$.

\par
  By $\gamma\in \mathcal{G}_i\subseteq K_i^\times$, we have $\gamma^{1+q^{\frac{d_i}{3}}+q^{\frac{2d_i}{3}}}=1$
and $\gamma^{q^{d_i}}=\gamma$, which implies
$\gamma^{q^{\frac{2d_i}{3}}+1}=\gamma^{q^{-\frac{d_i}{3}}}$.
Then from
$$X^{q^{\frac{d_i}{3}}+1}\equiv -(\gamma^{q^{\frac{2d_i}{3}}}X+\gamma^{-1})\ ({\rm mod} \ \phi_{i,\gamma}(X)),$$
we deduce that
\begin{eqnarray*}
X^{q^{\frac{2d_i}{3}}+q^{\frac{d_i}{3}}+1}
&\equiv&(X^{q^{\frac{d_i}{3}}+1})^{q^{\frac{d_i}{3}}}X\\
  &\equiv& -(\gamma^{q^{\frac{2d_i}{3}}}X+\gamma^{-1})^{q^{\frac{d_i}{3}}}X\\
  &\equiv & -(\gamma^{q^{d_i}}X^{q^{\frac{d_i}{3}}+1}+\gamma^{q^{-\frac{d_i}{3}}}X)\\
  &\equiv& -\gamma(X^{q^{\frac{d_i}{3}}+1}+\gamma^{q^{\frac{2d_i}{3}}}X)\\
  &\equiv & -\gamma\gamma^{-1}=1 \ ({\rm mod} \ \phi_{i,\gamma}(X)),
\end{eqnarray*}
which implies that $\phi_{i,\gamma}(X)$ is a factor of $X^{q^{\frac{2d_i}{3}}+q^{\frac{d_i}{3}}+1}-1$.
Hence
$\alpha^{q^{\frac{2d_i}{3}}+q^{\frac{d_i}{3}}+1}=1$ for any root $\alpha$ of $\phi_{i,\gamma}(X)$.
Therefore, all roots of $\phi_{i,\gamma}(X)$ are contained in $\mathcal{G}_i$.

\par
  The reciprocal polynomial
of $\psi_{i,\gamma}(X)$ is equal to
$$\psi_{i,\gamma}^\ast(X)=X^{q^{\frac{d_i}{3}}+1}+\gamma^{-1}X+\gamma^{q^{\frac{d_i}{3}}}.$$
Then by $X^{q^{\frac{d_i}{3}}+1}\equiv -(\gamma^{-1}X+\gamma^{q^{\frac{d_i}{3}}})$ (mod $\psi_{i,\gamma}^\ast(X)$) and
$\gamma^{-q^{\frac{d_i}{3}}-1}=\gamma^{q^{\frac{2d_i}{3}}}$, it follows that
\begin{eqnarray*}
X^{q^{\frac{2d_i}{3}}+q^{\frac{d_i}{3}}+1}
&\equiv&(X^{q^{\frac{d_i}{3}}+1})^{q^{\frac{d_i}{3}}}X\\
 &\equiv& -(\gamma^{-1}X+\gamma^{q^{\frac{d_i}{3}}})^{q^{\frac{d_i}{3}}}X\\
 &\equiv&-(\gamma^{-q^{\frac{d_i}{3}}}X^{q^{\frac{d_i}{3}}+1}+\gamma^{q^{\frac{2d_i}{3}}}X)\\
 &\equiv&-(-\gamma^{-q^{\frac{d_i}{3}}}(\gamma^{-1}X+\gamma^{q^{\frac{d_i}{3}}})+\gamma^{q^{\frac{2d_i}{3}}}X)\\
 &\equiv& 1 \ ({\rm mod} \ \psi_{i,\gamma}^\ast(X)),
\end{eqnarray*}
which implies that $\psi_{i,\gamma}^\ast(x)$ is a factor of $X^{q^{\frac{2d_i}{3}}+q^{\frac{d_i}{3}}+1}-1$. Since
$X^{q^{\frac{2d_i}{3}}+q^{\frac{d_i}{3}}+1}-1$ is self-reciprocal, we conclude that $\psi_{i,\gamma}(X)$ is also a factor of $X^{q^{\frac{2d_i}{3}}+q^{\frac{d_i}{3}}+1}-1$. Therefore, all roots of $\psi_{i,\gamma}(X)$ are contained in $\mathcal{G}_i$.
\end{IEEEproof}

\vskip 3mm
  \begin{theorem}\label{th3.6}
  (i) Let $\alpha,\beta\in \mathcal{G}_i$. Then $C_{i,1,\beta}\subset C_{i,2,\alpha}$
if and only of $\alpha$ and $\beta$ satisfying the following equation
\begin{equation}
\label{eq5}
\alpha\theta_i(\alpha)\beta+\alpha\beta\theta_i^2(\beta)+1=0.
\end{equation}

\par
  (ii) For any $\beta\in \mathcal{G}_i$, there are
$q^{\frac{d_i}{3}}+1$ codes $C_{i,2,\alpha}$ containing $C_{i,1,\beta}$ where $\alpha$
is given by one of the following two case:

\par
   $\alpha$ is a root of the polynomial $\phi_{i,\beta}(X)$ if $\theta_i(a(x))=a(x)^{q^{\frac{d_i}{3}}}$ for all $a(x)\in K_i$;

\par
   $\alpha$ is a root of the polynomial $\psi_{i,\beta}(X)$ if $\theta_i(a(x))=a(x)^{q^{\frac{2d_i}{3}}}$ for all $a(x)\in K_i$.

\par
  (iii) For any $\alpha\in \mathcal{G}_i$, there are
$q^{\frac{d_i}{3}}+1$ codes $C_{i,1,\beta}$ contained in $C_{i,2,\alpha}$ where $\beta$
is given by one of the following two case:

\par
   $\beta$ is a root of the
polynomial $\psi_{i,\alpha}(X)$ if $\theta_i(a(x))=a(x)^{q^{\frac{d_i}{3}}}$ for all $a(x)\in K_i$;

\par
   $\beta$ is a root of the
polynomial $\phi_{i,\alpha}(X)$ if $\theta_i(a(x))=a(x)^{q^{\frac{2d_i}{3}}}$ for all $a(x)\in K_i$.
\end{theorem}

\vskip 3mm
  \begin{IEEEproof}
   (i) It is clear that $C_{i,1,\beta}\subset C_{i,2,\alpha}$ if and only if
$-\alpha+y$ is a right divisor of $\beta^{-1}+\theta_i^2(\beta)y+y^2$.

\par
   Dividing $\beta^{-1}+\theta_i^2(\beta)y+y^2$ by $y-\alpha$ from right hand in the skew polynomial ring $K_i[y;\theta_i]$, we have
\begin{eqnarray*}
&&\beta^{-1}+\theta_i^2(\beta)y+y^2\\
&=&(\theta_i(\alpha)+\theta_i^2(\beta))(-\alpha+y)+\beta^{-1}+\alpha\theta_i(\alpha)+\alpha\theta_i^2(\beta).
\end{eqnarray*}
Hence $-\alpha+y$ is a right divisor of $\beta^{-1}+\theta_i^2(\beta)y+y^2$ if and only if
$\beta^{-1}+\alpha\theta_i(\alpha)+\alpha\theta_i^2(\beta)=0$, which is equivalent to that
$\alpha$ and $\beta$ satisfy (\ref{eq5}).

\par
  (ii) Let $\alpha\in \mathcal{G}_i$. We have one of the following two cases:

\par
   When $\theta_i(a(x))=a(x)^{q^{\frac{d_i}{3}}}$ for all $a(x)\in K_i$, it is clear that
$$\alpha\theta_i(\alpha)\beta+\alpha\beta\theta_i^2(\beta)+1=\beta\phi_{i,\beta}(\alpha).$$
From this and by (i), we deduce that $C_{i,1,\beta}\subset C_{i,2,\alpha}$
if and only if $\phi_{i,\beta}(\alpha)=0$. Furthermore, by Lemma \ref{lm3.5} we know that the
$\phi_{i,\beta}(x)$ has exactly $q^{\frac{d_i}{3}}+1$ roots in $\mathcal{G}_i$.

\par
   When $\theta_i(a(x))=a(x)^{q^{\frac{2d_i}{3}}}$ for all $a(x)\in K_i$, by
$\alpha\theta_i(\alpha)=\alpha^{1+q^{\frac{2d_i}{3}}}=q^{-q^{\frac{d_i}{3}}}$ we have
$$\alpha\theta_i(\alpha)\beta+\alpha\beta\theta_i^2(\beta)+1=\alpha^{1+q^{\frac{2d_i}{3}}}\beta\psi_{i,\beta}(\alpha).$$
From this and by (i), we deduce that $C_{i,1,\beta}\subset C_{i,2,\alpha}$
if and only if $\psi_{i,\beta}(\alpha)=0$. Furthermore, by Lemma \ref{lm3.5} we know that the
polynomial $\psi_{i,\beta}(x)$ has exactly $q^{\frac{d_i}{3}}+1$ roots in $\mathcal{G}_i$.

\par
  (iii) By (\ref{eq5}) and Lemma \ref{lm3.5}, it can be proved similarly as that of (ii). Here, we omit the proof.
\end{IEEEproof}

%%%%%%%%%%%%%%%%%%%%%%%%%%%%%%%%%%%%%%%%%%%%%%%%%%%%%%%%%%%%%%%%%%%%%%%%%%%%%%%%%%%%%%%%%

%%%%%%%%%%%%%%%%%%%%%%%%%%%%%%%%%%%%%%%%%%%%%%%%%%%%%%%%%%%%%%%%%%%%%%%%%%%%%%%%%%%%%%%%%%%%%%
\section{The dual code of any left $G_{(m,3,r)}$-code}
\noindent
 In this section, we give the dual code of any left $G_{(m,3,r)}$-code over $\mathbb{F}_q$
and determine all self-orthogonal left $G_{(m,3,r)}$-codes.

\par
   As in \cite{jitman2013abelian}, the \textit{Euclidian inner product} in $\mathbb{F}_{q}[G_{(m,3,r)}]$ is defined as follows.
For $\xi=\sum_{i=0}^{m-1}\sum_{j=0}^2a_{i,j}x^iy^j$ and $\eta=\sum_{i=0}^{m-1}\sum_{j=0}^2b_{i,j}x^iy^j$ in $\mathbb{F}_{q}[G_{(m,3,r)}]$, we set
$$[\xi,\eta]_E=\sum_{i=0}^{m-1}\sum_{j=0}^2a_{i,j}b_{i,j}\in \mathbb{F}_q.$$
The \textit{Euclidian dual code} of a left $G_{(m,3,r)}$-code $\mathcal{C}$ over $\mathbb{F}_{q}$ is defined by
$$\mathcal{C}^{\bot_E}=\{\xi\in \mathbb{F}_{q}[G_{(m,3,r)}]\mid [\xi,\eta]_E=0, \ \forall \eta\in \mathcal{C}\}.$$
$\mathcal{C}$ is said to be \textit{self-orthogonal} if $\mathcal{C}\subseteq \mathcal{C}^{\bot_E}$.

\par
  For any $\xi(y)=\sum_{i=0}^{m-1}\sum_{j=0}^2a_{i,j}x^iy^j\in \mathbb{F}_{q}[G_{(m,3,r)}]$,
as in \cite{jitman2013abelian} we define the \textit{conjugation} $\mu$ on $\mathbb{F}_{q}[G_{(m,3,r)}]$ by
$$\mu(\xi(y))=\sum_{i=0}^{m-1}\sum_{j=0}^2a_{i,j}y^{-j}x^{-i} \ \ \ \ \ \ \ \ \ \ \ \ \ \ \ \ \ \ \ \ \ \ \ \ \ \ \  $$
\begin{equation}
\label{eq6}
=a_0(x^{-1})+y^2a_1(x^{-1})+ya_2(x^{-1}),
\end{equation}
where $a_j(x)=\sum_{i=0}^{m-1}a_{i,j}x^{i}$ and $a_j(x^{-1})=a_{0,j}+\sum_{i=1}^{m-1}a_{i,j}x^{m-i}$
for $j=0,1,2$.
It can be verify easily that
$$\mu(\xi+\eta)=\mu(\xi)+\mu(\eta) \ {\rm and} \ \mu(\xi\eta)=\mu(\eta)\mu(\xi)$$
for all $\xi,\eta\in \mathbb{F}_{q}[G_{(m,3,r)}]$.
Moreover, we have the following

\vskip 3mm
\begin{lemma}\label{lm4.1}
 (i) The map $\mu$ defined by $(6)$ is an $\mathbb{F}_{q}$-algebra anti-automorphism of $\mathbb{F}_{q}[G_{(m,3,r)}]$ satisfying $\mu^{-1}=\mu$.

\par
  (ii) For any $\xi,\eta\in \mathbb{F}_{q}[G_{(m,3,r)}]$, we have
$[\xi,\eta]_E=0$ if $\xi\cdot \mu(\eta)=0$ in the ring $\mathbb{F}_{q}[G_{(m,3,r)}]$.

\par
  (iii) Let $\mathcal{C}$ be a left $G_{(m,3,r)}$-code over $\mathbb{F}_q$ and $\mathcal{B}$
a right ideal of $\mathbb{F}_{q}[G_{(m,3,r)}]$. Then
$$\mu(\mathcal{B})\subseteq \mathcal{C}^{\perp_E} \  {\rm if} \
\mathcal{C}\cdot \mathcal{B}=\{0\} \  {\rm in} \ \mathbb{F}_{q}[G_{(m,3,r)}].$$
\end{lemma}

\vskip 3mm
\begin{IEEEproof}
  (ii) For $\xi=\sum_{i=0}^{m-1}\sum_{j=0}^2a_{i,j}x^iy^j$ and $\eta=\sum_{i=0}^{m-1}\sum_{j=0}^2b_{i,j}x^iy^j$ in $\mathbb{F}_{q}[G_{(m,3,r)}]$, by ${\rm ord}(x)=m$ and ${\rm ord}(y)=3$ we deduce that
\begin{eqnarray*}
\xi\cdot\mu(\eta)&=&(\sum_{i=0}^{m-1}\sum_{j=0}^2a_{i,j}x^iy^j)(\sum_{k=0}^{m-1}\sum_{l=0}^2b_{k,l}y^{3-l}x^{m-k})\\
  &=&[\xi,\eta]_E+\sum_{0\leq i\leq m-1, 0\leq j\leq 2,i+j\neq 0}c_{i,j}x^iy^j
\end{eqnarray*}
for some $c_{i,j}\in \mathbb{F}_q$. Hence $[\xi,\eta]_E=0$ if $\xi\cdot \mu(\eta)=0$ in the ring $\mathbb{F}_{q}[G_{(m,3,r)}]$.

\par
  (iii) For any $\beta\in \mathcal{B}$ and $\eta\in \mathcal{C}$, by $\mathcal{C}\cdot \mathcal{B}=\{0\}$ we have $\eta\beta=0$,
which implies $\mu(\beta)\cdot\mu(\eta)=0$ by (i), and so $[\mu(\beta),\eta]_E=0$ by (ii). From this we deduce
that $\mu(\beta)\in\mathcal{C}^{\bot_E}$. Therefore, $\mu(\mathcal{B})\subseteq \mathcal{C}^{\bot_E}$.
\end{IEEEproof}

\vskip 3mm
\par
  By the identification of $\mathbb{F}_q[G_{(m,3,r)}]$ with ${\cal A}[y;\theta]/\langle y^3-1\rangle$, we see that
$\mathcal{A}=\mathbb{F}_q[x]/\langle x^m-1\rangle$ is a subring of $\mathbb{F}_{q}[G_{(m,3,r)}]$. In the following,
we consider the restriction of $\mu$ on $\mathcal{A}$. In order to simplify the notation, we still denote this
restriction by $\mu$. Obviously, we have
$$\mu(a(x))=a(x^{-1})=\sum_{i=0}^{m-1}a_ix^{-i}=a_0+\sum_{i=1}^{m-1}a_ix^{m-i}$$
for all $a(x)=\sum_{i=0}^{m-1}a_ix^{i}\in \mathcal{A}$. It is clear that $\mu$ is an
$\mathbb{F}_{q}$-algebra automorphism of $\mathcal{A}$ satisfying $\mu^{-1}=\mu$.

\par
   Using the notations of Section 2, we know that $J(i)=J_{k_i}^{(q)}$, $0\leq i\leq s+t$,
are the all distinct $q$-cyclotomic cosets modulo $m$. By Lemma \ref{lm2.2}, we have one of the following two cases:

\par
   $\diamondsuit$ $0\leq i\leq s$. In this case, we have $\theta(j)=rj\equiv j$ (mod $m$) for all $j\in J(i)=J_{k_i}^{(q)}$.
Then it is clear that $-J(i)=\{-j\mid j\in J(i)\}$ (mod $m$) is a $q$-cyclotomic coset modulo $m$ satisfying
$\theta(-j)=-rj\equiv -j$ (mod $m$) for all $j\in J(i)$. Hence $-J(i)$ is also a $q$-cyclotomic coset modulo $m$ satisfying
Condition (I) in Lemma \ref{lm2.2}. Therefore, there is a unique integer $i^\prime$, $0\leq i^\prime\leq s$, such
that $-J(i)=J(i^\prime)$.

\par
   $\diamondsuit$ $s+1\leq i\leq s+t$. In this case, we have $\theta(j)\in J(i)$ and $\theta(j)\not\equiv j$ (mod $m$) for all $j\in J(i)=J_{k_i}^{(q)}$. Then it is clear that $-J(i)$ is a $q$-cyclotomic coset modulo $m$ satisfying
$\theta(-j)=-\theta(j)\in -J(i)$ and $\theta(-j)\not\equiv -j$ (mod $m$) for all $j\in J(i)$. Hence $-J(i)$ is also a $q$-cyclotomic coset modulo $m$ satisfying
Condition (II) in Lemma \ref{lm2.2}. Therefore, there is a unique integer $i^\prime$, $s+1\leq i^\prime\leq s+t$, such
that $-J(i)=J(i^\prime)$.

\par
  We also use $\mu$ to denote this map $i\mapsto i^\prime$, i.e., $\mu(i)=i^\prime$. Whether $\mu$ denotes
the automorphism of $\mathcal{A}$
or this map on the set $\{0,1,\ldots,s+t\}$ is determined by context. The next lemma
shows the compatibility of the two uses of $\mu$.

\vskip 3mm
\begin{lemma}\label{lm4.2}
  Using the notations above, the following assertions hold.

\par
  (i) $\mu$ is a permutation on $\{0,1,\ldots,s+t\}$ satisfying $\mu^{-1}=\mu$, $\mu(0)=0$,
$1\leq \mu(i)\leq s$ for all $1\leq i\leq s$ and $s+1\leq \mu(i)\leq s+t$ for all $s+1\leq i\leq s+t$.

\par
  (ii) After a rearrangement of $J(0),J(1)\ldots,J(s+t)$, there are nonnegative integers
$s_1,s_2,t_1,t_2$ satisfying the following conditions:

\par
  $\bullet$ $s=s_1+2s_2$, $\mu(i)=i$ for all $1\leq i\leq s_1$,
$\mu(i)=i+s_2$ and $\mu(i+s_2)=i$ for all $s_1+1\leq i\leq s_1+s_2$;

\par
  $\bullet$ $t=t_1+2t_2$, $\mu(i)=i$ for all $s+1\leq i\leq s+t_1$,
$\mu(i)=i+t_2$ and $\mu(i+t_2)=i$ for all $s+t_1+1\leq i\leq s+t_1+t_2$.

\par
  (iii) $\mu(\varepsilon_i(x))=\varepsilon_{\mu(i)}(x)$
and $\mu(\mathcal{A}_i)=\mathcal{A}_{\mu(i)}$  for all
$i=0,1,\ldots,s+t$.

\par
  (iv) Let $\mu$ be the map defined by $\mu(a(x))=a(x^{-1})=a(x^{m-1})$ $({\rm mod} \ f_{\mu(i)}(x))$
for all $a(x)\in K_i=\mathbb{F}_q[x]/\langle f_i(x)\rangle$. Then $\mu$ is an $\mathbb{F}_q$-algebra isomorphism from
$K_i$ onto $K_{\mu(i)}=\mathbb{F}_q[x]/\langle f_{\mu(i)}(x)\rangle$ satisfying $\mu\theta_i=\theta_{\mu(i)}\mu$.

\par
  (v) Let $0\leq i\leq s+t$. Using the notations of Theorem \ref{th2.4}(ii), the $\mathbb{F}_{q}$-algebra anti-automorphism $\mu$
of $\mathbb{F}_{q}[G_{(m,3,r)}]$ induces an $\mathbb{F}_q$-algebra anti-isomorphism $\varphi_{\mu(i)}^{-1}\mu\varphi_i$ from
$\mathcal{R}_i$ onto $\mathcal{R}_{\mu(i)}$. We denote this anti-isomorphism by $\mu$ as well. Then
for any $\alpha(y)=a(x)+b(x)y+c(x)y^2\in \mathcal{R}_i$ where $a(x),b(x),c(x)\in K_i$, we have
\begin{equation}
\label{eq7}
\mu(\alpha(y))=a(x^{-1})+y^2b(x^{-1})+yc(x^{-1}).
\end{equation}
\end{lemma}

\vskip 1mm
\begin{IEEEproof}
   (i) follows from the definition of the map $\mu$, and (ii) follows from (i).

\par
  (iii) It is clear that $\mu(\varepsilon_i(x))=\frac{1}{m}\sum_{l=0}^{m-1}(\sum_{j\in J(i)}\zeta^{-jl})x^{-l}$
by (\ref{eq3}) in Section 2. From this, by $\mathbb{Z}_m=-\mathbb{Z}_m$ and $J(\mu(i))=-J(i)=\{-j\mid j\in J(i)\}$ we deduce that
\begin{eqnarray*}
\mu(\varepsilon_i(x))&=&\frac{1}{m}\sum_{l=0}^{m-1}(\sum_{j\in J(i)}\zeta^{-(-j)(-l)})x^{-l}\\
 &=&\frac{1}{m}\sum_{k=0}^{m-1}(\sum_{j^\prime\in J(\mu(i))}\zeta^{-j^\prime k})x^{k}\\
  &=&\varepsilon_{\mu(i)}(x).
\end{eqnarray*}
Hence $\mu(\mathcal{A}_i)=\mu(\mathcal{A}\varepsilon_i(x))=\mu(\mathcal{A})\mu(\varepsilon_i(x))
=\mathcal{A}\varepsilon_{\mu(i)}(x)=\mathcal{A}_{\mu(i)}$ by Lemma \ref{lm2.3}(ii).

\par
  (iv) By (iii), we know that $\mu$ induces an $\mathbb{F}_q$-algebra
isomorphism from $\mathcal{A}_i$ onto $\mathcal{A}_{\mu(i)}$. Then by Lemma \ref{lm2.3}(iii), we see that $\varphi_{\mu(i)}^{-1}\mu\varphi_i$
is an $\mathbb{F}_q$-algebra
isomorphism from $K_i$ onto $K_{\mu(i)}$. For any $a(x)\in K_i$, by Equation (\ref{eq3}) in Section 2 we have $\varepsilon_{\mu(i)}(x)\equiv 1$ (mod $f_{\mu(i)}(x)$), which implies
\begin{eqnarray*}
(\varphi_{\mu(i)}^{-1}\mu\varphi_i)(a(x))&=&\varphi_{\mu(i)}^{-1}\mu(\varepsilon_i(x)a(x))\\
  &=&\varphi_{\mu(i)}^{-1}(\varepsilon_{\mu(i)}(x)a(x^{-1}))\\
  &=&a(x^{-1}) \ ({\rm mod} \ f_{\mu(i)}(x)).
\end{eqnarray*}
Since we denote $\varphi_{\mu(i)}^{-1}\mu\varphi_i$ by
$\mu$ as well, the map $\mu: a(x)\mapsto a(x^{-1})$ (mod $f_{\mu(i)}(x)$) is an $\mathbb{F}_q$-algebra
isomorphism from $K_i$ onto $K_{\mu(i)}$. Moreover,
for any $a(x)\in K_i$ by Lemma \ref{lm2.3}(v) and $a(x^{-1})\in K_{\mu(i)}$ it follows that
\begin{eqnarray*}
(\mu\theta_i)(a(x))&=&\mu(a(x^r))=a(x^{-r})=\theta_{\mu(i)}(a(x^{-1}))\\
&=&(\theta_{\mu(i)}\mu)(a(x)).
\end{eqnarray*}
Hence $\mu\theta_i=\theta_{\mu(i)}\mu$.

\par
  (v) By (iii) and Theorem \ref{th2.4}(ii), we have the following commutative diagram
form ring isomorphisms:
$$\left.\begin{array}{ccc}\ \ \ \ \ \ \ \ \ \mathcal{R}_i  & \stackrel{\varphi_i}{\longrightarrow} &
\mathcal{A}_i[y;\theta|_{\mathcal{A}_i}]/\langle \varepsilon_i(x)(y^3-1)\rangle
 \cr {\small \varphi_{\mu(i)}^{-1}\mu\varphi_i}\downarrow  &    &  \ \ \ \ \downarrow \mu \cr
\ \ \ \ \ \ \ \ \ \mathcal{R}_{\mu(i)} & \stackrel{\varphi_{\mu(i)}}{\longrightarrow} & \mathcal{A}_{\mu(i)}[y;\theta|_{\mathcal{A}_{\mu(i)}}]/\langle \varepsilon_{\mu(i)}(x)(y^3-1)\rangle\end{array}\right.$$
As we write $\varphi_{\mu(i)}^{-1}\mu\varphi_i$ by $\mu$, for any $a(x),b(x),c(x)\in K_i$
by the identification of $\mathbb{F}_q[G_{(m,3,r)}]$ with ${\cal A}[y;\theta]/\langle y^3-1\rangle$, $\varepsilon_i(x^{-1})=\mu(\varepsilon_i(x))=\varepsilon_{\mu(i)}(x)$, Equation (\ref{eq6}) and
$y\varepsilon_{\mu(i)}(x)=\theta(\varepsilon_{\mu(i)}(x))y=\varepsilon_{\mu(i)}(x)y$, we deduce that
\begin{eqnarray*}
&&\mu(a(x)+b(x)y+c(x)y^2)\\
&=&(\varphi_{\mu(i)}^{-1}\mu)(\varphi_i(a(x)+b(x)y+c(x)y^2))\\
  &=&\varphi_{\mu(i)}^{-1}\left(\mu(\varepsilon_i(x)a(x)+\varepsilon_i(x)b(x)y+\varepsilon_i(x)c(x)y^2)\right)\\
  &=&\varphi_{\mu(i)}^{-1}(\varepsilon_i(x^{-1})a(x^{-1})+y^2\varepsilon_i(x^{-1})b(x^{-1})\\
  &&+y\varepsilon_i(x^{-1})c(x^{-1}))\\
  &=&\varphi_{\mu(i)}^{-1}\left(\varepsilon_{\mu(i)}(x)\left(a(x^{-1})+y^2b(x^{-1})+yc(x^{-1})\right)\right)\\
  &=&a(x^{-1})+y^2b(x^{-1})+yc(x^{-1})
\end{eqnarray*}
by (iv).
\end{IEEEproof}

\vskip 3mm
\begin{corollary}\label{co4.3}
 For any $\alpha(x)\in \mathcal{G}_i$, we denote
$$\widehat{\alpha}(x)=(\alpha(x^{m-1}))^{q^{\frac{2d_i}{3}}+q^{\frac{d_i}{3}}} \ ({\rm mod} \ f_{\mu(i)}(x)).$$
Using the notations of Lemma \ref{lm4.2}(iv), we have that $\widehat{\alpha}(x)=(\alpha(x^{-1}))^{-1}=(\mu(\alpha(x)))^{-1}
\in \mathcal{G}_{\mu(i)}$, $\alpha(x^{-1})\widehat{\alpha}(x)=1$ and $\alpha(x)=(\widehat{\alpha}(x^{-1}))^{-1}$.
\end{corollary}

\vskip 3mm
\begin{IEEEproof} As $\alpha(x)\in \mathcal{G}_i$, we see that $\alpha(x)$ is an element
of $K_i=\mathbb{F}_q[x]/\langle f_i(x)\rangle$ satisfying
$(\alpha(x))^{q^{\frac{2d_i}{3}}+q^{\frac{d_i}{3}}+1}=1$. By Lemma \ref{lm4.2}(iv), we know that $\mu$ is an $\mathbb{F}_q$-algebra isomorphism from $K_i$ onto
$K_{\mu(i)}=\mathbb{F}_q[x]/\langle f_{\mu(i)}(x)\rangle$. Hence $(\mu(\alpha(x)))^{-1}\in K_{\mu(i)}$
and $(\mu(\alpha(x)))^{q^{\frac{2d_i}{3}}+q^{\frac{d_i}{3}}+1}=1$ in
$K_{\mu(i)}$, which implies $\mu(\alpha(x))\in \mathcal{G}_{\mu(i)}$, and so $(\mu(\alpha(x))^{-1}\in \mathcal{G}_{\mu(i)}$. Finally, by $f_{\mu(i)}(x)|(x^m-1)$ it follows that
\begin{eqnarray*}
\widehat{\alpha}(x)&=&(\alpha(x^{-1}))^{q^{\frac{2d_i}{3}}+q^{\frac{d_i}{3}}}=(\mu(\alpha(x)))^{q^{\frac{2d_i}{3}}+q^{\frac{d_i}{3}}}\\
 &=&(\mu(\alpha(x)))^{-1}
\end{eqnarray*}
in $K_{\mu(i)}$. Then $\alpha(x)=(\mu(\widehat{\alpha}(x)))^{-1}=(\widehat{\alpha}(x^{-1}))^{-1}$.
\end{IEEEproof}

\vskip 3mm
\par
   For any integer $i$, $0\leq i\leq s+t$, and $g(y),h(y)\in \mathcal{R}_i=K_i[y;\theta_i]/\langle y^3-1\rangle$,
in the following we define
$$g(y)\sim_l h(y) \ {\rm if} \ g(y)=\alpha h(y) \ {\rm for} \ {\rm some} \ \alpha\in \mathcal{R}_i^\times,$$
where $\mathcal{R}_i^\times$ is the set of invertible elements in $\mathcal{R}_i$.
It is clear that $\mathcal{R}_ig(y)=\mathcal{R}_ih(y)$ if $g(y)=\alpha h(y)$.

\vskip 3mm
  \begin{lemma}\label{lm4.4} For any integer $i$, $0\leq i\leq s+t$, we have the following conclusions:

\par
  (i) $\mu(y^2+y+1)\sim_l y^2+y+1$, $\mu(y^3-1)\sim_l y^3-1$ and $\mu((y-1)^j)\sim_l (y-1)^j$ for all $j=0,1,2$.

\par
  (ii) Let $0\leq i\leq s$ and $q^{d_i}\equiv 1$ (mod $3$). Then $\mu(y-\omega_i(x))\sim_l y-\omega_i(x^{-1})^2$
and $\mu(y-\omega_i(x)^2)\sim_l y-\omega_i(x^{-1})$ in $\mathcal{R}_{\mu(i)}$.

\par
 (iii) Let $s+1\leq i\leq s+t$ and $\alpha(x)\in \mathcal{G}_i$. Then
$$\mu(-\alpha(x)+y)\sim_l -\theta_{\mu(i)}(\widehat{\alpha}(x))+y,$$
\begin{eqnarray*}
&&\mu\left(\alpha(x)^{-1}+\theta_i^2(\alpha(x)) y+y^2\right)\\
&\sim_l&\left(\theta_{\mu(i)}^2(\widehat{\alpha}(x))\right)^{-1}
+\theta_{\mu(i)}^2\left(\theta_{\mu(i)}^2(\widehat{\alpha}(x))\right)y+y^2.
\end{eqnarray*}
\end{lemma}

\vskip 3mm
\begin{IEEEproof}
 (i) By Equation (\ref{eq7}) and $y^3=1$, it follows that
$\mu(y-1)=y^{2}-1=(-y^2)(y-1)$ where $-y^2\in \mathcal{R}_i^\times$. The other conclusion
can be verified similarly.

\par
  (ii) Since $\omega_i(x)^3=1$ and $\mu$ is a ring isomorphism from $\mathcal{R}_i$ onto
$\mathcal{R}_{\mu(i)}$ by Lemma \ref{lm4.2}(v), it follows that $\omega_i(x^{-1})^3=(\mu(\omega_i(x)))^3=1$.
As $-\omega_i(x^{-1})y^2\in \mathcal{R}_{\mu(i)}^\times$, we have
\begin{eqnarray*}
\mu(y-\omega_i(x))&=&-\omega_i(x^{-1})+y^2\\
  &=&(-\omega_i(x^{-1})y^2)(y-\omega_i(x^{-1})^2)\\
  &\sim_l&y-\omega_i(x^{-1})^2.
\end{eqnarray*}
Similarly, one can verify that $\mu(y-\omega_i(x)^2)\sim_l y-\omega_i(x^{-1})$.

\par
  (iii) By (\ref{eq7}), Lemma \ref{lm4.2}(v) and Corollary \ref{co4.3} we have
\begin{eqnarray*}
\mu(-\alpha(x)+y)
 &=&-\alpha(x^{-1})+y^2\\
 &=&-\mu(\alpha(x))\left(1-\widehat{\alpha}(x)y^2\right)\\
 &=&-\mu(\alpha(x))\left(1-y^2\theta_{\mu(i)}(\widehat{\alpha}(x))\right)\\
 &=&-\mu(\alpha(x))y^2\cdot y\left(1-y^2\theta_{\mu(i)}(\widehat{\alpha}(x))\right)\\
 &=&-\mu(\alpha(x))y^2\left(-\theta_{\mu(i)}(\widehat{\alpha}(x))+y\right),
\end{eqnarray*}
where $-\mu(\alpha(x))y^2\in \mathcal{R}_{\mu(i)}^\times$ and $\theta_{\mu(i)}(\widehat{\alpha}(x))\in \mathcal{G}_{\mu(i)}$ by Corollary \ref{co4.3}. Similarly, we have
\begin{eqnarray*}
&&\mu\left(\alpha(x)^{-1}+\theta_i^2(\alpha(x)) y+y^2\right)\\
&=&\alpha(x^{-1})^{-1}+y^2\theta_{\mu(i)}^2(\alpha(x^{-1})) +y\\
  &=&y^2\cdot y\left(\widehat{\alpha}(x)+y^2\theta_{\mu(i)}^2(\widehat{\alpha}(x)^{-1})+y\right)\\
  &=&y^2\left(\theta_{\mu(i)}^2(\widehat{\alpha}(x)^{-1}) +\theta_{\mu(i)}(\widehat{\alpha}(x))y+y^2\right)
\end{eqnarray*}
where $y^2\in \mathcal{R}_{\mu(i)}^\times$, $\theta_{\mu(i)}^2(\widehat{\alpha}(x)^{-1})=\left(\theta_{\mu(i)}^2(\widehat{\alpha}(x))\right)^{-1}\in \mathcal{G}_{\mu(i)}$
by Corollary 4.3 and $\theta_{\mu(i)}(\widehat{\alpha}(x))=\theta_{\mu(i)}^2\left(\theta_{\mu(i)}^2(\widehat{\alpha}(x))\right)$.
\end{IEEEproof}

\vskip 3mm
\par
   Now, we give the dual code of any left $G_{(m,3,r)}$-code over $\mathbb{F}_q$ by
the following theorem.

\vskip 3mm
\begin{theorem}\label{th4.5}
Let ${\cal C}=\oplus_{i=0}^{s+t}({\cal A}_i\Box_{\varphi_i}C_i)$ be
a left $G_{(m,3,r)}$-code over $\mathbb{F}_q$, where $C_i$ is a left ideal of the ring ${\cal R}_i=K_i[y;\theta_i]/\langle y^3-1\rangle$
given by Theorems \ref{th3.1} and \ref{th3.3}. Then the dual code ${\cal C}^{\bot_E}$ of ${\cal C}$ is also a left $G_{(m,3,r)}$-code over $\mathbb{F}_q$. Precisely, we have
$${\cal C}^{\bot_E}=\bigoplus_{i=0}^{s+t}({\cal A}_i\Box_{\varphi_i}D_i),$$
where $D_i$ is a left ideal of ${\cal R}_i$ given by one of the following cases:

\par
 (i) Let $0\leq i\leq s$. Then $\mathcal{R}_i=K_i[y]/\langle y^3-1\rangle$ and
$D_i$ is given by one of the following subcases:

\par
 (i-1)  Let $q\equiv 0$ $({\rm mod} \ 3)$. Then $D_{\mu(i)}=\mathcal{R}_{\mu(i)}\frac{(y-1)^3}{g(y)}$,
if $C_i=\mathcal{R}_{i}g(y)$ where $g(y)\in\{1,y-1,(y-1)^2,(y-1)^3\}$.

\par
 (i-2)  Let $q^{d_i}\equiv 2$ $({\rm mod} \ 3)$. Then $D_{\mu(i)}=\mathcal{R}_{\mu(i)}\frac{y^3-1}{g(y)}$,
if $C_i=\mathcal{R}_{i}g(y)$ where $g(y)\in\{1,y-1,y^2+y+1,y^3-1\}$.

\par
 (i-3)  Let $q^{d_i}\equiv 1$ $({\rm mod} \ 3)$. Then $D_{\mu(i)}=\mathcal{R}_{\mu(i)}\vartheta(y)$ if $C_i=\mathcal{R}_{i}g(y)$,
where the pair $(g(y),\vartheta(y))$ of polynomials is given by the following table:

\vskip 1mm
\begin{center}
\begin{tabular}{l|l}\hline
$g(y)$ & $\vartheta(y)$ (${\rm mod} \ f_{\mu(i)}(x)$)\\ \hline
$1$ & $y^3-1$,\\
$y-1$ & $y^2+y+1$\\
$y-\omega_i(x)$ & $(y-1)(y-\omega_i(x^{-1}))$ \\
$y-\omega_i(x)^2$ & $(y-1)(y-\omega_i(x^{-1})^2)$ \\
$y^2+y+1$ & $y-1$ \\
$(y-1)(y-\omega_i(x)^2)$ & $y-\omega_i(x^{-1})^2$ \\
$(y-1)(y-\omega_i(x))$ & $y-\omega_i(x^{-1})$ \\
$y^3-1$ & $1$ \\ \hline
\end{tabular}
\end{center}

\vskip 1mm \par
  (ii) Let $s+1\leq i\leq s+t$. Then $\mathcal{R}_i=K_i[y;\theta_i]/\langle y^3-1\rangle$ and
$D_i$ is given by one of the following subcases:

\par
  (ii-1) $D_{\mu(i)}={\cal R}_{\mu(i)}\left(-\theta_{\mu(i)}(\widehat{\alpha}(x))+y\right)=C_{\mu(i),2,\theta_{\mu(i)}(\widehat{\alpha})}$ and a generator matrix
of $D_{\mu(i)}$ is given by
$$G_{D_{\mu(i)}}=\left(\begin{array}{ccc}-\theta_{\mu(i)}(\widehat{\alpha}(x)) & 1 & 0 \cr
                                        0 & -\theta_{\mu(i)}\left(\theta_{\mu(i)}(\widehat{\alpha}(x))\right) & 1 \end{array}\right)$$
as a linear code over $K_{\mu(i)}$ of length $3$, if
$C_i=C_{i,1,\alpha}=\mathcal{R}_{i}\left(\alpha(x)^{-1}+\theta_i^2(\alpha(x))y+y^2\right)$ where $\alpha(x)\in \mathcal{G}_i$.

\par
  (ii-2) $D_{\mu(i)}={\cal R}_{\mu(i)}(\theta_{\mu(i)}^2(\widehat{\alpha}(x)^{-1})
   +\theta_{\mu(i)}(\widehat{\alpha}(x))y+y^2 )=C_{\mu(i),1,\theta_{\mu(i)}^2(\widehat{\alpha})}$ and a generator matrix
of $D_{\mu(i)}$ is given by
$$G_{D_{\mu(i)}}=\left(\left(\theta_{\mu(i)}^2(\widehat{\alpha}(x))\right)^{-1},
\theta_{\mu(i)}^2\left(\theta_{\mu(i)}^2(\widehat{\alpha}(x))\right),1\right)$$
as a linear code over $K_{\mu(i)}$ of length $3$, if $C_i=C_{i,2,\alpha}=\mathcal{R}_{i}(-\alpha(x)+y)$ where $\alpha(x)\in \mathcal{G}_i$.

\par
 (ii-3) $D_{\mu(i)}=\{0\}$ if $C_i={\cal R}_i$; $D_{\mu(i)}={\cal R}_{\mu(i)}$ if $C_i=\{0\}$.
\end{theorem}

\vskip 3mm
\begin{IEEEproof}
 Let $K_i=\mathbb{F}_q[x]/\langle f_i(x)\rangle$, $\mathcal{R}_i=K_i[y;\theta_i]/\langle y^3-1\rangle$
and $B_i$ be an right ideal of the ring $\mathcal{R}_i=K_i[y;\theta_i]/\langle y^3-1\rangle$ given by one of the following two cases:

\par
   (A) Let $0\leq i\leq s$. Then $\theta_i={\rm id}_{K_i}$, $\mathcal{R}_i=K_i[y]/\langle y^3-1\rangle$
and $B_i$ is given by one of the following three subcases.

\par
   (A-1) Let $q\equiv 0$ (mod $3$). Then $B_i=\frac{y^3-1}{g(y)}\mathcal{R}_i$ if
$C_i=\mathcal{R}_ig(y)$ where $g(y)\in\{1, y-1, (y-1)^2, y^3-1\}$.

\par
   (A-2) Let $q^{d_i}\equiv 2$ (mod $3$). Then $B_i=\frac{y^3-1}{g(y)}\mathcal{R}_i$ if
$C_i=\mathcal{R}_ig(y)$ where $g(y)\in\{1, y-1, y^2+y+1, y^3-1\}$.

\par
   (A-3) Let $q^{d_i}\equiv 1$ (mod $3$). Then $B_i=\frac{y^3-1}{g(y)}\mathcal{R}_i$ if
$C_i=\mathcal{R}_ig(y)$ where $g(y)\in\{1, y-1, y-\omega_i(x), y-\omega_i(x)^2, y^2+y+1, ( y-1)(y-\omega_i(x)^2), ( y-1)(y-\omega_i(x)),y^3-1\}$.

\par
   (B) Let $s+1\leq i\leq s+t$. Then $B_i$ is given by one of the following four subcases.

\par
   (B-1) $B_i=(-\alpha(x)+y)\mathcal{R}_i$,
if $C_i=\mathcal{R}_i(\alpha(x)^{-1}+\theta_i^2(\alpha(x))y+y^2)$ where $\alpha(x)\in\mathcal{G}_i$.

\par
   (B-2) $B_i=\left(\alpha(x)^{-1}+\theta_i^2(\alpha(x))y+y^2\right)\mathcal{R}_i$,
if $C_i=\mathcal{R}_i(-\alpha(x)+y)$ where $\alpha(x)\in\mathcal{G}_i$.

\par
   (B-3) $B_i=\{0\}$, if $C_i=\mathcal{R}_i$; $B_i=\mathcal{R}_i$, if $C_i=\{0\}$.

\noindent
  By Theorems \ref{th3.1} and \ref{th3.3}, Lemma \ref{lm3.2}(v) and direct calculations, one can easily
verify that
\begin{equation}
\label{eq8}
C_i\cdot B_i=\{0\} \ {\rm in} \ \mathcal{R}_i, \ i=0,1,\ldots,s+t.
\end{equation}

\par
   For any integer $0\leq i\leq s+t$, let $D_{\mu(i)}=\mu(B_i)$. By Lemma \ref{lm4.2}(v) we see that $D_{\mu(i)}$ is
a left ideal of $\mathcal{R}_{\mu(i)}$. Let
$$\mathcal{D}=\sum_{i=0}^{s+t}\mathcal{A}_i\Box_{\varphi_i}D_i
=\bigoplus_{i=0}^{s+t}\mathcal{A}_{\mu(i)}\Box_{\varphi_{\mu(i)}}D_{\mu(i)}.$$
Then by Theorem \ref{th2.4}(iii), we conclude that $\mathcal{D}$ is a left $G_{(m,3,r)}$-code over $\mathbb{F}_q$.

\par
  $\diamondsuit$  First, we give the clear expression of $D_{\mu(i)}=\mu(B_i)$.

\par
   For the trivial case: $B_i=\mathcal{R}_i$ or $B_i=\{0\}$, the conclusion follows from Lemma 4.2(v) immediately.
Then we only need to consider the nontrivial cases in (A) and (B).

\par
  In the case of (A-1), $B_i=\frac{y^3-1}{g(y)}\mathcal{R}_i$. If $g(y)=(y-1)^2$, then
$B_i=(y-1)\mathcal{R}_i$.  By Lemma \ref{lm4.2}(v) and Lemma \ref{lm4.4}(i), we have
$D_{\mu(i)}=\mu(B_i)=\mathcal{R}_{\mu(i)}\mu(y-1)=\mathcal{R}_{\mu(i)}(y-1)$.

\par
  Similarly, one can easily prove that the other conclusions in (i-1) and all conclusions in (i-2) hold from (A-1) and (A-2).

\par
  In the case of (A-3), $B_i=\frac{y^3-1}{g(y)}\mathcal{R}_i$. If $g(y)=(y-1)(y-\omega(x)^2)$,
by Lemma \ref{lm4.2}(v) and Lemma \ref{lm4.4}(ii), we have
\begin{eqnarray*}
D_{\mu(i)}&=&\mu((y-\omega_i(x))\mathcal{R}_i)=\mu(\mathcal{R}_i)\mu((y-\omega_i(x))\\
  &=&\mathcal{R}_{\mu(i)}(y-\omega_i(x^{-1})^2).
\end{eqnarray*}

\par
  Similarly, one can easily prove that the other conclusions in (i-3) hold from (A-3).

\par
   In the case of (B-1), by Lemma \ref{lm4.2}(v) and Lemma \ref{lm4.4}(iii) we have
\begin{eqnarray*}
D_{\mu(i)}&=&\mu\left((-\alpha(x)+y)\mathcal{R}_i\right)
   =\mu(\mathcal{R}_i)\mu(-\alpha(x)+y)\\
 &=&{\cal R}_{\mu(i)}\left(-\theta_{\mu(i)}(\widehat{\alpha}(x))+y\right)\\
 &=&C_{\mu(i),2,\theta_{\mu(i)}(\widehat{\alpha})}
\end{eqnarray*}
Hence the conclusion (ii-1) holds by Theorem \ref{th3.3}(ii).

\par
  Similarly, in the case of (B-2) we have
\begin{eqnarray*}
&&D_{\mu(i)}
 =\mu(\mathcal{R}_i)\mu\left(\alpha(x)^{-1}+\theta_i^2(\alpha(x)) y+y^2\right)\\
&=&{\cal R}_{\mu(i)}\left((\theta_{\mu(i)}^2(\widehat{\alpha}(x)))^{-1}
+\theta_{\mu(i)}^2(\theta_{\mu(i)}^2(\widehat{\alpha}(x)))y+y^2\right)\\
&=&C_{\mu(i),1,\theta_{\mu(i)}^2(\widehat{\alpha})}.
\end{eqnarray*}
Hence the conclusion (ii-2) holds by Theorem \ref{th3.3}(iii).

\par
  $\diamondsuit$ Then we prove that $|\mathcal{C}||\mathcal{D}|=|\mathbb{F}_q|^{3m}$. For any $0\leq i\leq s+t$, by Theorems \ref{th3.1}, \ref{th3.3} and direct calculations we deduce that
$|C_i||D_{\mu(i)}|=|K_i|^3=|\mathcal{R}_i|$. From this and by Theorem \ref{th2.4} (i)--(iii), we obtain
\begin{eqnarray*}
|\mathcal{C}||\mathcal{D}|&=&(\prod_{i=0}^{s+t}|C_i|)(\prod_{i=0}^{s+t}|D_{\mu(i)}|)=\prod_{i=0}^{s+t}|C_i||D_{\mu(i)}|\\
  &=&\prod_{i=0}^{s+t}|\mathcal{R}_i|=\prod_{i=0}^{s+t}|\mathcal{A}_i[y;\theta|_{\mathcal{A}_i}]/\langle \varepsilon_i(x)(y^3-1)\rangle|\\
   &=&|\mathcal{A}[y;\theta]/\langle y^3-1\rangle|=|\mathbb{F}_q[G_{(m,3,r)}]|\\
   &=&|\mathbb{F}_q|^{3m}.
\end{eqnarray*}

\par
  $\diamondsuit$ We claim that $\mathcal{D}\subseteq \mathcal{C}^{\bot_E}$. In fact, let $\xi\in \mathcal{D}$ and $\eta\in \mathcal{C}$. Then for each integer $i$, $0\leq i\leq s+t$,
there exist $\alpha_i\in C_i$ and $\beta_i\in D_i$ such that $\xi=\sum_{i=0}^{s+t}\varepsilon_i(x)\alpha_i$
and $\eta=\sum_{i=0}^{s+t}\varepsilon_i(x)\beta_i$, where $C_i$ and $D_i$ are left ideals of $\mathcal{R}_i$ given by
(i)--(ii) and $\varepsilon_i(x)\alpha_i, \varepsilon_i(x)\beta_i\in \mathcal{A}_i[y;\theta|_{\mathcal{A}_i}]/\langle \varepsilon_i(x)(y^3-1)\rangle$. By Lemma \ref{lm2.3}(iv), we see that $\varepsilon_i(x)$ is the multiplicative identity of $\mathcal{A}_i[y;\theta|_{\mathcal{A}_i}]/\langle \varepsilon_i(x)(y^3-1)\rangle$. Since
$\varepsilon_i(x)\varepsilon_j(x)=0$ for all $0\leq i\neq j\leq s+t$, we have $\varepsilon_i(x)\varepsilon_{\mu(j)}(x)=0$
if $i\neq\mu(j)$, i.e., $j\neq\mu(i)$. Hence
\begin{eqnarray*}
\xi\cdot\mu(\eta)&=&(\sum_{i=0}^{s+t}\varepsilon_i(x)\alpha_i)(\sum_{i=0}^{s+t}\mu(\varepsilon_i(x)\beta_i)\\
 &=&\left(\sum_{i=0}^{s+t}(\varepsilon_i(x)\alpha_i)\varepsilon_i(x)\right)\\
  &&\cdot\left(\sum_{i=0}^{s+t}\varepsilon_{\mu(i)}(x)(\mu(\beta_i)\varepsilon_{\mu(i)}(x))\right)\\
 &=&\sum_{i,j=0}^{s+t}(\varepsilon_i(x)\alpha_i)
 \varepsilon_i(x)\varepsilon_{\mu(j)}(x)(\mu(\beta_j)\varepsilon_{\mu(j)}(x))\\
 &=&\sum_{i=0}^{s+t}(\varepsilon_i(x)\alpha_i)
 \varepsilon_i(x)(\mu(\beta_{\mu(i)})\varepsilon_{i}(x))\\
 &=&\sum_{i=0}^{s+t}\varepsilon_i(x)(\alpha_i\mu(\beta_{\mu(i)})).
\end{eqnarray*}
Since $\beta_{\mu(i)}\in D_{\mu(i)}$, by Lemma \ref{lm4.2}(i) we see that
$$\mu(\beta_{\mu(i)})\in \mu(D_{\mu(i)})=\mu(\mu(B_i))=B_i.$$
From this and by (\ref{eq8}), we deduce that $\alpha_i\mu(\beta_{\mu(i)})=0$ for all $i=0,1,\ldots,s+t$,
which implies $\xi\cdot\mu(\eta)=0$, and so $[\xi,\eta]_E=0$ by Lemma \ref{lm4.1}(ii). Therefore,
$\mathcal{D}\subseteq \mathcal{C}^{\bot_E}$.

\par
   As stated above, we conclude that $\mathcal{D}=\mathcal{C}^{\bot_E}$ since both $\mathcal{C}$ and $\mathcal{D}$ are linear codes over $\mathbb{F}_q$ of length $3m$.
\end{IEEEproof}

\vskip 3mm
\par
   Finally, we determine self-orthogonal left $G_{(m,3,r)}$-codes.

\vskip 3mm
\begin{theorem}\label{th4.6}
  All distinct self-orthogonal left $G_{(m,3,r)}$-codes over $\mathbb{F}_q$
are given by the following
$$\mathcal{C}=\bigoplus_{i=0}^{s+t}\mathcal{A}_i\Box_{\varphi_i}C_i=\bigoplus_{i=0}^{s+t}\{\varepsilon_i(x)\xi\mid \xi\in C_i\} \ ({\rm mod} \ x^m-1),$$
where $C_i$ is an left ideal of $\mathcal{R}_i=K_i[y;\theta_i]/\langle y^3-1\rangle$ given by one of the
following four cases:

\par
  (i) $0\leq i\leq s_1$. In this case, $C_i$ is given by one of the following three subcases.

\par
  (i-1) If $q\equiv 0$ (mod $3$), $C_i=\{0\}$ or $C_i=\mathcal{R}_i(y-1)^2$.

\par
  (i-2) If $q^{d_i}\equiv 2$ (mod $3$), $C_i=\{0\}$.

\par
  (i-3) Let $q^{d_i}\equiv 1$ (mod $3$). If $\omega_i(x^{-1})\equiv \omega_i(x)$ (mod $f_i(x)$), then
$C_i=\{0\}$, $C_i=\mathcal{R}_i(y-1)(y-\omega_i(x))$ or $C_i=\mathcal{R}_i(y-1)(y-\omega_i(x)^2)$.
Otherwise,
$C_i=\{0\}$.

\par
  (ii) $s_1+1\leq i\leq s_1+s_2$. In this case, $C_i=\mathcal{R}_ig(y)$,
$C_{i+s_2}=\mathcal{R}_{i+s_2}\vartheta(y)$ and the pair $(g(y),\vartheta(y))$ of polynomials is given by one of the following three subcases.

\par
  (ii-1) Let $q\equiv 0$ (mod $3$). There are $10$ pairs $(g(y),\vartheta(y))$:

\vskip 1mm  \begin{center}
\begin{tabular}{l|l}\hline
$g(y)$ & $\vartheta(y)$ \\ \hline
$y^3-1$ & $y^3-1,(y-1)^2,y-1,1$ \\
$(y-1)^2$ & $y^3-1,(y-1)^2,y-1$ \\
$y-1$ & $y^3-1,(y-1)^2$ \\
$1$ & $y^3-1$ \\ \hline
\end{tabular}
\end{center}

\vskip 1mm \par
  (ii-2) Let $q^{d_i}\equiv 2$ (mod $3$). There are $9$ pairs $(g(y),\vartheta(y))$:

\vskip 1mm   \begin{center}
\begin{tabular}{l|l}\hline
$g(y)$ & $\vartheta(y)$ \\ \hline
$y^3-1$ & $y^3-1,y^2+y+1,y-1,1$ \\
$y^2+y+1$ & $y^3-1,y-1$ \\
$y-1$ & $y^3-1,y^2+y+1$ \\
$1$ & $y^3-1$ \\ \hline
\end{tabular}
\end{center}

\vskip 1mm \par
  (ii-3) Let $q^{d_i}\equiv 1$ (mod $3$). There are $27$ pairs $(g(y),\vartheta(y))$:

\vskip 1mm
{\small  \begin{center}
\begin{tabular}{l|l}\hline
$g(y)$ & $\vartheta(y)$ (${\rm mod}$ $f_{\mu(i)}(x)$)\\ \hline
$y^3-1$ & $1,y^2+y+1,(y-1)(y-\omega_i(x^{-1}))$,\\
        & $(y-1)(y-\omega_i(x^{-1})^2),y-1$,\\
        & $y-\omega_i(x^{-1})^2, y-\omega_i(x^{-1}),y^3-1$ \\ \hline
$y^2+y+1$ & $y^3-1,(y-1)(y-\omega_i(x^{-1}))$,\\
          & $(y-1)(y-\omega_i(x^{-1})^2),y-1$ \\ \hline
$(y-1)(y-\omega(x))$ & $y^3-1,(y-1)(y-\omega_i(x^{-1}))$,\\
          & $y^2+y+1,y-\omega_i(x^{-1})$ \\  \hline
$(y-1)(y-\omega(x)^2)$ & $y^3-1,(y-1)(y-\omega_i(x^{-1})^2)$,\\
          & $y^2+y+1,y-\omega_i(x^{-1})^2$ \\  \hline
$y-1$ & $y^3-1,y^2+y+1$ \\
$y-\omega(x)$ & $y^3-1,(y-1)(y-\omega_i(x^{-1}))$ \\
$y-\omega(x)^2$ & $y^3-1,(y-1)(y-\omega_i(x^{-1})^2)$ \\
$1$ & $y^3-1$ \\ \hline
\end{tabular}
\end{center} }

\vskip 1mm \par
  (iii) $s+1\leq i\leq s+t_1$. In this case, $C_i=\{0\}$ or
$C_i=C_{i,1,\alpha}=\mathcal{R}_i(\alpha(x)^{-1}+\theta_i^2(\alpha(x))y+y^2)$ where
$\alpha=\alpha(x)\in \mathcal{G}_i$ satisfying
$$\theta_{i}(\widehat{\alpha})\theta_{i}^2(\widehat{\alpha})\alpha+\theta_{i}(\widehat{\alpha})\alpha\theta_i^2(\alpha)+1=0.$$

\par
  (iv) $s+t_1+1\leq i\leq s+t_1+t_2$. In this case, there are exactly
$10+8q^{\frac{d_i}{3}}+8q^{\frac{2d_i}{3}}+q^{d_i}$ pairs $(C_i,C_{i+t_2})$ given by
one of the following four subcases:

\par
  (iv-1) $4+2q^{\frac{d_i}{3}}+2q^{\frac{2d_i}{3}}$ pairs: $(\{0\},C_{i+t_2})$, where
$C_{i+t_2}$ is any left ideal of $\mathcal{R}_{i+t_2}$ listed by Theorem \ref{th3.3}.

\par
  (iv-2) $(1+q^{\frac{d_i}{3}}+q^{\frac{2d_i}{3}})(3+q^{\frac{d_i}{3}})$ pairs:
$(C_{i,1,\alpha},C_{i+t_2})$, where $\alpha=\alpha(x)\in \mathcal{G}_i$ and $C_{i+t_2}$
is one of the following $3+q^{\frac{d_i}{3}}$ left ideals of $\mathcal{R}_{i+t_2}$:

\par
  $\bullet$ $C_{i+t_2}=\{0\}$.

\par
  $\bullet$ $C_{i+t_2}=C_{i+t_2,1,\beta}$ where $\beta=\theta_{i+t_2}^2(\gamma(x^{-1})^{-1})$ (mod $f_{i+t_2}(x)$)
and $\gamma(x)\in \mathcal{G}_i$ satisfying the following conditions:

\par
  $\phi_{i,\alpha}(\gamma(x))=0$, if $\theta_i(a(x))=a(x)^{q^{\frac{d_i}{3}}}$
for all $a(x)\in K_i$;

\par
  $\psi_{i,\alpha}(\gamma(x))=0$, if $\theta_i(a(x))=a(x)^{q^{2\frac{d_i}{3}}}$
for all $a(x)\in K_i$.

\par
  $\bullet$ $C_{i+t_2}=C_{i+t_2,2,\beta}$ where $\beta=\theta_{i+t_2}(\alpha(x^{-1})^{-1})$ (mod $f_{i+t_2}(x)$).

\par
  (iv-3) $2(1+q^{\frac{d_i}{3}}+q^{\frac{2d_i}{3}})$ pairs:
$(C_{i,2,\alpha},C_{i+t_2})$, where $\alpha=\alpha(x)\in \mathcal{G}_i$ and $C_{i+t_2}$
is one of the following $2$ left ideals of $\mathcal{R}_{i+t_2}$:

\par
  $\bullet$ $C_{i+t_2}=\{0\}$.

\par
  $\bullet$ $C_{i+t_2}=C_{i+t_2,1,\beta}$ where $\beta=\theta_{i+t_2}^2(\alpha(x^{-1})^{-1})$ (mod $f_{i+t_2}(x)$).

\par
  (iv-1) $1$ pair:
$(\mathcal{R}_i,\{0\})$.
\end{theorem}

\vskip 3mm
\begin{IEEEproof}
  By Theorem \ref{th4.5} and its proof we have $\mathcal{C}^{\bot_E}=\oplus_{i=0}^{s+t}\mathcal{A}_i\Box_{\varphi_i}D_i$,
where $D_i=\mu(B_{\mu(i)})$. From this and by Theorem \ref{th2.4}, we deduce that $\mathcal{C}$ is a self-orthogonal left $G_{(m,3,r)}$-code over $\mathbb{F}_q$
if and only if $C_i\subseteq D_i=\mu(B_{\mu(i)})$ for all $i=0,1,\ldots,s+t$.

\par
  By Equation (\ref{eq8}) and the proof of Theorem \ref{th4.5}, it follows
that $C_{\mu(i)}\cdot B_{\mu(i)}=\{0\}$ and
$$|C_{\mu(i)}||B_{\mu(i)}|=|C_{\mu(i)}||\mu(B_{\mu(i)})|=|C_{\mu(i)}||D_{i}|=|\mathcal{R}_{\mu(i)}|,$$
which implies that $C_{\mu(i)}$ is the
annihilating left ideal of $B_{\mu(i)}$ in $\mathcal{R}_{\mu(i)}$, i.e.,
\begin{equation}
\label{eq9}
C_{\mu(i)}={\rm Ann}^{(L)}_{\mathcal{R}_{\mu(i)}}(B_{\mu(i)})
\end{equation}
where ${\rm Ann}^{(L)}_{\mathcal{R}_{\mu(i)}}(B_{\mu(i)})=\{\xi\in \mathcal{R}_{\mu(i)}\mid \xi\eta=0, \ \forall \eta\in B_{\mu(i)}\}$.

\par
  Since $\mu$ is an $\mathbb{F}_q$-algebra
anti-isomorphism from $\mathcal{R}_i$ onto $\mathcal{R}_{\mu(i)}$, by
$C_i\cdot B_i=\{0\}$ and $|C_i||B_i|=|\mathcal{R}_i|$ we have
$\mu(B_i)\cdot \mu(C_i)=\{0\}$ and $|\mu(B_i)||\mu(C_i)|=|\mathcal{R}_{\mu(i)}|$, which
implies
$$ {\rm Ann}^{(L)}_{\mathcal{R}_{\mu(i)}}(\mu(C_i))=\mu(B_i)=D_{\mu(i)}.$$
From this, by $D_i=\mu(B_{\mu(i)})$ and (\ref{eq9}), we deduce that
\begin{eqnarray*}
C_i\subseteq D_i &\Longleftrightarrow& B_{\mu(i)}\supseteq \mu(C_i) \\
 &\Longleftrightarrow& {\rm Ann}^{(L)}_{\mathcal{R}_{\mu(i)}}(B_{\mu(i)})\subseteq {\rm Ann}^{(L)}_{\mathcal{R}_{\mu(i)}}(\mu(C_i))\\
 &\Longleftrightarrow& C_{\mu(i)}\subseteq D_{\mu(i)}
\end{eqnarray*}
for all $i=0,1,\ldots,s+t$.

\par
  $\diamondsuit$ Let $0\leq i\leq s$. Then $\theta_i={\rm id}_{K_i}$ and that $\mathcal{R}_i=K_i[y]/\langle y^3-1\rangle$ is a commutative ring. By Theorem \ref{th4.5} and its proof there is a unique pair $(g(y),h(y))$
of monic factors $g(y),h(y)$ of $y^3-1$ in $K_i[y]$ such that

\par
  $C_i=\mathcal{R}_ig(y)$, $D_i=\mathcal{R}_i\frac{y^3-1}{h(y)}=\mu(B_{\mu(i)})$ with $B_{\mu(i)}=\mathcal{R}_{\mu(i)}\mu(\frac{y^3-1}{h(y)})$,
and
\begin{equation}
\label{eq10}
C_{\mu(i)}={\rm Ann}^{(L)}_{\mathcal{R}_{\mu(i)}}(B_{\mu(i)})=\mathcal{R}_{\mu(i)}\mu(h(y)).
\end{equation}
Hence
\begin{equation}
\label{eq11}
C_i\subseteq D_i \Longleftrightarrow  \frac{y^3-1}{h(y)}\mid g(y)\Longleftrightarrow (y^3-1)\mid g(y)h(y).
\end{equation}
By Lemma \ref{lm4.2}(ii) and Theorems \ref{th3.1}, we have one of
the following two cases.

\par
  (i) When $0\leq i\leq s_1$, $\mu(i)=i$, which implies $\mu(h(y))\sim_l g(y)$ by (\ref{eq10}), and so $\mu(g(y))\sim_l h(y)$. Form this and by
(\ref{eq11}) we deduce that $C_i\subseteq D_i$ if and only if $(y^3-1)\mid g(y)\mu(g(x))$. Then the conclusions follow
from Lemma \ref{lm4.4} (i) and (ii).

\par
  (ii) Let $s_1+1\leq i\leq s_1+s_2$, $\mu(i)=i+s_2$. By Lemma \ref{lm4.4} (i) and (ii), we see that
for each monic factor $h(y)$ of $y^3-1$ in $K_i[y]$ there is a unique monic factor $\vartheta(y)$ of $y^3-1$ in $K_{\mu(i)}[y]$
such that $\mu(h(y))\sim_l \vartheta(y)$, which implies $C_{i+s_2}=\mathcal{R}_{\mu(i)}\vartheta(y)$ by (\ref{eq10}). Then the conclusions follow
from (\ref{eq11}), Lemma \ref{lm4.4} (i) and (ii) immediately.

\par
  $\diamondsuit$ Let $s+1\leq i\leq s+t$. Then $\mathcal{R}_i=K_i[y;\theta_i]/\langle y^3-1\rangle$ is a noncommutative ring.
By Theorem \ref{th4.5}(ii), the pair $(C_i,D_{\mu(i)})$ is given by one of the following cases:

\par
  $\diamond$ $C_i=\{0\}$ and $D_{\mu(i)}=\mathcal{R}_{\mu(i)}$, or $C_i=\mathcal{R}_{i}$ and $D_{\mu(i)}=\{0\}$;

\par
  $\diamond$ $C_i=C_{i,1,\alpha}$ and $D_{\mu(i)}=C_{\mu(i),2,\theta_{\mu(i)}(\widehat{\alpha})}$, where
$\alpha\in\mathcal{G}_i$;

\par
  $\diamond$ $C_i=C_{i,2,\alpha}$ and $D_{\mu(i)}=C_{\mu(i),1,\theta_{\mu(i)}^2(\widehat{\alpha})}$, where
$\alpha\in\mathcal{G}_i$.

\noindent
From these, we deduce that
\begin{equation}
\label{eq12}
{\rm dim}_{K_i}(C_i)+{\rm dim}_{K_{\mu(i)}}(C_{\mu(i)})=3.
\end{equation}
  Then by Lemma \ref{lm4.2}(ii), we have one and only one of the following two cases.

\par
  (iii) Let $s+1\leq i\leq s+t_1$. Then $\mu(i)=i$. In this case, we deduce that
the condition $C_i\subseteq D_i$ if and only if $C_i=\{0\}$ or $C_i=C_{i,1,\alpha}$ where $\alpha=\alpha(x)\in \mathcal{G}_i$ satisfying
$C_{i,1,\alpha}\subset C_{i,2,\theta_{i}(\widehat{\alpha})}$. By Theorem \ref{th3.6}(i), the condition
$C_{i,1,\alpha}\subset C_{i,2,\theta_{i}(\widehat{\alpha})}$ is equivalent to that
$\theta_{i}(\widehat{\alpha})\theta_{i}(\theta_{i}(\widehat{\alpha}))\alpha+\theta_{i}(\widehat{\alpha})\alpha\theta_i^2(\alpha)+1=0$,
i.e.,
$\theta_{i}(\widehat{\alpha})\theta_{i}^2(\widehat{\alpha})\alpha+\theta_{i}(\widehat{\alpha})\alpha\theta_i^2(\alpha)+1=0.$

\par
  (iv) Let $s+t_1+1\leq i\leq s+t_1+t_2$. Then $\mu(i)=i+t_2$. By Theorem \ref{th4.5}(ii), we have one of the following four
situations:

\par
  (iv-1) Let $C_i=\{0\}$. Then $D_{i+t_2}=\mathcal{R}_{\mu(i)}$. In this case,
$C_{i+t_2}\subseteq D_{i+t_2}$ for any left ideal $C_{i+t_2}$ of $\mathcal{R}_{\mu(i)}$.

\par
  By Theorem \ref{th3.3}, the number of pairs $(\{0\}, C_{i+t_2})$ is equal to
$4+2q^{\frac{d_i}{3}}+2q^{\frac{2d_i}{3}}$.

\par
  (iv-2) Let $C_i=C_{i,1,\alpha}$ where $\alpha=\alpha(x)\in \mathcal{G}_i$.
Then ${\rm dim}_{K_{i+t_i}}(D_{i+t_2})=2$ by (\ref{eq12}) and
$$D_{i+t_2}=C_{i+t_2,2,\theta_{i+t_2}(\widehat{\alpha})}$$
by Theorem \ref{th4.5}(ii-1). Hence ${\rm dim}_{K_{i+t_i}}(C_{i+t_2})\leq 2$ if $C_{i+t_2}\subseteq D_{i+t_2}$. Then we have
one of the following three cases.

\par
  $\triangleright$ It is obvious that $C_{i+t_2}=\{0\}$ satisfying $C_{i+t_2}\subseteq D_{i+t_2}$.

\par
  $\triangleright$ Let $C_{i+t_2}=C_{i+t_2,1,\beta}$ where $\beta=\beta(x)\in \mathcal{G}_{i+t_2}$.
By Lemma \ref{lm4.2}(ii), we have $\mu(i+t_2)=i$, which implies
$D_i=D_{\mu(i+t_2)}=C_{i,2,\theta_i(\widehat{\beta})}$ by Theorem \ref{th4.5}(ii-1).
From this and by
Theorem \ref{th3.6}(ii), we deduce that $C_i\subset D_i$ if and only if $\theta_i(\widehat{\beta})$ satisfies the following
conditions:

\par
  $\phi_{i,\alpha}(\theta_i(\widehat{\beta}))=0$, if $\theta_i(a(x))=a(x)^{q^{\frac{d_i}{3}}}$ for all $a(x)\in K_i$.

\par
  $\psi_{i,\alpha}(\theta_i(\widehat{\beta}))=0$, if $\theta_i(a(x))=a(x)^{q^{\frac{2d_i}{3}}}$ for all $a(x)\in K_i$.

\noindent
  We denote $\gamma=\gamma(x)=\theta_i(\widehat{\beta}(x))$. Then
$\gamma=\theta_i(\mu(\beta(x))^{-1})\in \mathcal{G}_i$ by Corollary \ref{co4.3}, which implies
$\mu(\beta(x))^{-1}=\theta_i^2(\gamma(x))$, and hence $\beta(x)=\theta_{i+t_2}^2(\gamma(x^{-1})^{-1})$ by Lemma \ref{lm4.2}(iv).
Moreover, by Lemma \ref{lm3.5} we know that both $\phi_{i,\alpha}(x)$ and $\psi_{i,\alpha}(x)$ have exactly $q^{\frac{d_i}{3}}+1$ roots in $\mathcal{G}_i$.

\par
  $\triangleright$ Let $C_{i+t_2}=C_{i+t_2,2,\beta}$ where $\beta=\beta(x)\in \mathcal{G}_{i+t_2}$.
As $\mu(i+t_2)=i$, we have
$D_i=D_{\mu(i+t_2)}=C_{i,1,\theta_i^2(\widehat{\beta})}$ by Theorem \ref{th4.5}(ii-2).
Hence $C_i=C_{i,1,\alpha}\subseteq D_i$ if and only if $\alpha=\theta_i^2(\widehat{\beta})=\theta_i^2(\beta(x^{-1})^{-1})$,
which is equivalent to that $\beta(x)=\theta_{i+t_2}(\alpha(x^{-1})^{-1})$  by Lemma \ref{lm4.2}(iv).

\par
  Therefore, the number of pairs $(C_{i,1,\alpha},C_{i+t_2})$ is equal to
$$(1+q^{\frac{d_i}{3}}+q^{\frac{2d_i}{3}})(3+q^{\frac{d_i}{3}}).$$

\par
  (iv-3) Let $C_i=C_{i,2,\alpha}$ where $\alpha=\alpha(x)\in \mathcal{G}_i$.
Then ${\rm dim}_{K_{i+t_i}}(D_{i+t_2})=1$ by (\ref{eq12}) and
$$D_{i+t_2}=C_{i+t_2,1,\theta_{i+t_2}^2(\widehat{\alpha})}$$
by Theorem \ref{th4.5}(ii-2). Hence ${\rm dim}_{K_{i+t_i}}(C_{i+t_2})\leq 1$ if $C_{i+t_2}\subseteq D_{i+t_2}$. Then we have
one of the following two cases.

\par
  $\triangleright$ $C_{i+t_2}=\{0\}$.

\par
  $\triangleright$ Let $C_{i+t_2}=C_{i+t_2,1,\beta}$ where $\beta=\beta(x)\in \mathcal{G}_{i+t_2}$. Then
As $\mu(i+t_2)=i$, we have
$D_i=D_{\mu(i+t_2)}=C_{i,2,\theta_i(\widehat{\beta})}$ by Theorem \ref{th4.5}(ii-1).
Hence $C_i=C_{i,2,\alpha}\subseteq D_i$ if and only if $\alpha=\theta_i(\widehat{\beta})=\theta_i^2(\beta(x^{-1})^{-1})$,
which is equivalent to that $\beta(x)=\theta_{i+t_2}^2(\alpha(x^{-1})^{-1})$  by Lemma \ref{lm4.2}(iv).

\par
  Therefore, the number of pairs $(C_{i,1,\alpha},C_{i+t_2})$ is equal to
$2(1+q^{\frac{d_i}{3}}+q^{\frac{2d_i}{3}}).$

\par
  (iv-4) Let $C_i=\mathcal{R}_i$. Then $D_{i+t_2}=\{0\}$. From this and by
$C_{i+t_2}\subseteq D_{i+t_2}$, we deduce $C_{i+t_2}=\{0\}$.

\par
  As stated above, we conclude that the number of pairs $(C_i,C_{i+t_2})$, where $s+t_1+1\leq i\leq s+t_1+t_2$,
is equal to $10+8q^{\frac{d_i}{3}}+8q^{\frac{2d_i}{3}}+q^{d_i}.$
\end{IEEEproof}

%%%%%%%%%%%%%%%%%%%%%%%%%%%%%%%%%%%%%%%%%%%%%%%%%%%%%%%%%%%%%%%%%%%%%%%%%%%%%%%%%%%%%%%%%

%%%%%%%%%%%%%%%%%%%%%%%%%%%%%%%%%%%%%%%%%%%%%%%%%%%%%%%%%%%%%%%%%%%%%%%%%%%%%%%%%%%%%%%%%%%%%%
\section{An Example}
\noindent
  We consider left $G_{(14,3,9)}$-codes over $\mathbb{F}_3$.
Obviously, $9^3=729\equiv 1$ (mod $14$). All distinct $3$-cyclotomic cosets modulo $14$ are the following:
$J^{(3)}_0=\{0\}$, $J^{(3)}_7=\{7\}$, $J^{(3)}_2=\{2,6,4,12,8,10\}$, $J^{(3)}_1=\{1,3,9,13,11,5\}$. It is clear that
$$\theta(7)=9\cdot 7\equiv 7, \ \theta(2)=9\cdot 2\equiv 4, \ \theta(1)=9 \ ({\rm mod 14}).$$
Using the notations
is Section 2, we have that $s=1$, $t=2$, $J(0)=J^{(3)}_0$, $J(1)=J^{(3)}_7$, $J(2)=J^{(3)}_2$ and $J(3)=J^{(3)}_1$. Hence $d_0=d_1=1$, $d_i=6$ and $\frac{d_i}{3}=2$ for $i=2,3$.

\par
  Obviously, $3\equiv 0$ (mod $3$). By Corollary \ref{co3.4}(i),
the number of left $G_{(14,3,9)}$-codes over $\mathbb{F}_3$ is equal to
$$4^2\prod_{i=2,3}(4+2\cdot 3^{\frac{d_i}{3}}+2\cdot 3^{\frac{2d_i}{3}})=16\cdot 184^2=541,696.$$

\par
  We have $x^{14}-1=f_0(x)f_1(x)f_2(x)f_3(x)$, where $f_0(x)=x-1$, $f_1(x)=x+1$, $f_2(x)=x^6+x^5+x^4+x^3+x^2+x+1$
and $f_3(x)=x^6+2x^5+x^4+2x^3+x^2+2x+1$. Then

\par
  $\bullet$ $K_0=\mathbb{F}_3[x]/\langle x-1\rangle=\mathbb{F}_3$ and $\mathcal{R}_0=K_0[y]/\langle y^3-1\rangle
=\mathbb{F}_3[y]/\langle (y-1)^3\rangle$. By Theorem \ref{th3.1}(ii), all distinct ideals of $\mathcal{R}_0$ are given by:
$C_0=\mathcal{R}_0g(y)$, where
$g(y)\in \{1,y-1,(y-1)^2,y^3-1\}.$

\par
  $\bullet$ $K_1=\mathbb{F}_3[x]/\langle x+1\rangle=\mathbb{F}_3$ and $\mathcal{R}_1=K_1[y]/\langle y^3-1\rangle
=\mathbb{F}_3[y]/\langle (y-1)^3\rangle$. By Theorem \ref{th3.1}(ii), all distinct ideals of $\mathcal{R}_1$ are given by:
$C_1=\mathcal{R}_1g(y)$, where
$g(y)\in \{1,y-1,(y-1)^2,y^3-1\}.$

\par
Moreover, $|\langle 0\rangle|=1$, $|\mathcal{R}_0|=|\mathcal{R}_1|=3^3=27$, $|\mathcal{R}_0(y-1)|=|\mathcal{R}_1(y-1)|=3^2=9$ and $|\mathcal{R}_0(y-1)^2|=|\mathcal{R}_1(y-1)^2|=3$.

\par
  $\bullet$ $K_2=\mathbb{F}_2[x]/\langle f_2(x)\rangle=\{\sum_{j=0}^{5}a_jx^j\mid a_j\in \mathbb{F}_3, \ j=0,1,\ldots,5\}$
and $\varrho_2(x)=1+x$ is an element of multiplicative order $1+3^2+3^4=91$ in $K_2$.
Hence
$$\mathcal{G}_2=\{(1+x)^\lambda\mid \lambda=0,1,\ldots,90\} \ ({\rm mod} \ f_2(x))$$
and $\mathcal{R}_2=K_2[y;\theta_2]/\langle y^3-1\rangle$ where $\theta_2$ is an $\mathbb{F}_3$-algebra automorphism of $K_2$ defined by:
$$\theta_2(a(x))=a(x^r)=a(x^9)=a(x)^9 \ ({\rm mod} \ f_2(x))$$
for all $a(x)\in K_2$. Then
$\theta_2^2(\varrho_2(x))=(1+x)^{9^2}=1+x^4.$

\par
 By Theorem \ref{th3.3}, all distinct left ideals $C_2$ of $\mathcal{R}_2$ are given by the following three cases:

\par
   (i) $C_2=\{0\}$ with $|\{0\}|=1$, and $C_2=\mathcal{R}_2$ with $|\mathcal{R}_2|=|K_2|^3=(3^6)^3=3^{18}=387420489$.

\par
   (ii) $C_2=C_{2,2,(1+x)^\lambda}=\mathcal{R}_2(-(1+x)^\lambda+y)$ with $|C_2|=(3^6)^2=3^{12}=531441$, $\lambda=0,1,2,\ldots,90$.

\par
   (iii) $C_2=C_{2,1,(1+x)^\lambda}=\mathcal{R}_2((1+x)^{91-\lambda}+(1+x^4)^\lambda y+y^2)$ with $|C_2|=3^6=729$, $\lambda=0,1,2,\ldots,90$.

\par
  $\bullet$ $K_3=\mathbb{F}_2[x]/\langle f_3(x)\rangle=\{\sum_{j=0}^5a_jx^j\mid a_j\in \mathbb{F}_3\}$.
We find that $\varrho_3(x)=1+2x$ is an element of multiplicative order $1+3^2+3^4=91$ in $K_3$.
Hence
$$\mathcal{G}_3=\{(1+2x)^\lambda\mid \lambda=0,1,\ldots,90\} \ ({\rm mod} \ f_3(x))$$
and $\mathcal{R}_3=K_3[y;\theta_2]/\langle y^3-1\rangle$ where $\theta_3$ is an $\mathbb{F}_3$-algebra automorphism of $K_3$ defined by:
$$\theta_3(a(x))=a(x^r)=a(x^9)=a(x)^9 \ ({\rm mod} \ f_3(x))$$
for all $a(x)\in K_3$. In particular, we have
$\theta_3^2(\varrho_2(x))=(1+x)^{9^2}=1+x^4.$

\par
 By Theorem \ref{th3.3}, all distinct left ideals $C_3$ of $\mathcal{R}_3$ are given by the following three cases:

\par
   (i) $\{0\}$ and $\mathcal{R}_3$, where $|\{0\}|=1$ and $|\mathcal{R}_3|=|K_3|^3=(3^6)^3=3^{18}=387420489$.

\par
   (ii) $C_3=C_{3,2,(1+2x)^\lambda}=\mathcal{R}_3(-(1+2x)^\lambda+y)$ with $|C_3|=(3^6)^2=3^{12}=531441$, $\lambda=0,1,2,\ldots,90$.

\par
   (iii) $C_3=C_{3,1,(1+x)^\lambda}=\mathcal{R}_3((1+2x)^{91-\lambda}+(1+x^4)^\lambda y+y^2)$ with $|C_3|=3^6=729$, $\lambda=0,1,2,\ldots,90$.

\par
 $\bullet$ All distinct $541696$ left $G$-codes over $\mathbb{F}_3$ are given by
$$\mathcal{C}=\bigoplus_{i=0}^3\mathcal{A}_i\Box_{\varphi_i}C_i=\sum_{i=0}^3\{\varepsilon_i(x)\xi_i\mid \xi_i\in C_i\}$$
(mod $ x^{14}-1$) by Theorem \ref{th2.4}, where

\par
 $\varepsilon_0(x)=2+2x+2x^2+2x^3+2x^4+2x^5+2x^6+2x^7+2x^8+2x^9+2x^{10}+2x^{11}+2x^{12}+2x^{13}$,

 $\varepsilon_1(x)=2+x+2x^2+x^3+2x^4+x^5+2x^6+x^7+2x^8+x^9+2x^{10}+x^{11}+2x^{12}+x^{13}$,

 $\varepsilon_2(x)=x+x^2+x^3+x^4+x^5+x^6+x^8+x^9+x^{10}+x^{11}+x^{12}+x^{13}$,

$\varepsilon_3(x)=2x+x^2+2x^3+x^4+2x^5+x^6+x^8+2x^9+x^{10}+2x^{11}+x^{12}+2x^{13}$,

\noindent
  and the number of codewords in $\mathcal{C}$ is equal to

\par
\begin{center}  $|\mathcal{C}|=|C_0||C_1||C_2||C_3|$.\end{center}

\par
   As $-J(i)=J(i)$ (mod $14$), we have $\mu(i)=i$ for all $i=0,1,2,3$. Using the notations of Lemma \ref{lm4.2}, we have $s=s_1=1$, $s_2=0$,
$t=t_1=2$ and $t_2=0$. Hence $\mu(\varepsilon_i(x))=\varepsilon_i(x)$ for all $i=0,1,2,3$.

\par
  $\bullet$ By Theorem \ref{th4.6}, all self-orthogonal left $G_{(14,3,9)}$-codes over $\mathbb{F}_3$ are given by:
$\mathcal{C}=\bigoplus_{i=0}^3\mathcal{A}_i\Box_{\varphi_i}C_i$, where

\par
  $\diamond$ $C_0=\{0\}$ or $C_0=\mathcal{R}_0(y-1)^2$.

\par
  $\diamond$ $C_1=\{0\}$ or $C_1=\mathcal{R}_1(y-1)^2$.

\par
  $\diamond$ $C_2=\{0\}$ or $C_2=C_{2,1,\alpha(x)}$ where $\alpha(x)=(1+x)^\lambda$
satisfying
$\theta_{i}(\widehat{\alpha})\theta_{i}^2(\widehat{\alpha})\alpha+\theta_{i}(\widehat{\alpha})\alpha\theta_i^2(\alpha)+1=0$, i.e.,
\begin{eqnarray*}
0&=&1+(1+x^{-1})^{-9\lambda-81\lambda}(1+x)^\lambda\\
   &&+(1+x^{-1})^{-9\lambda}(1+x)^{\lambda+81\lambda}\\
 &=&1+x^{90\lambda}(1+x)^{-89\lambda}+x^{9\lambda}(1+x)^{73\lambda}
\end{eqnarray*}
in $K_2$ ($0\leq\lambda\leq 90$). Since $x^{14}=1$ and $(1+x)^{91}=1$ in $K_2$, the above condition is equivalent to
$$1+x^{6\lambda}(1+x)^{2\lambda}+x^{9\lambda}(1+x)^{73\lambda}\equiv 0 \ ({\rm mod} \ f_2(x)).$$
Precisely, we have
$$\lambda=0,7,8,11,13,20,21,24,26,33,34,37,39,46,47,$$
\begin{equation}\label{eq13}
\ \ \ \ \ \ \ \ \ \ 50,52,59,60,63,65,72,73,76,78,85,86,89.
\end{equation}

\par
  $\diamond$ $C_3=\{0\}$ or $C_3=C_{3,1,(1+2x)^\lambda}$ where $\lambda$ is given by (\ref{eq13}).

\vskip 2mm \par
  Therefore, the number of self-orthogonal left $G_{(14,3,9)}$-codes over $\mathbb{F}_3$ is equal to $2\cdot 2\cdot 29\cdot29=3364$.

\par
  For example, we have
$21$ self-orthogonal left $G_{(14,3,9)}$-codes over $\mathbb{F}_3$:
$\mathcal{C}=\{\varepsilon_2(x)\xi\mid \xi\in C_{2,1,(1+x)^\lambda}\}$ where $\lambda=7,8,11,20,21,24,33,34,37,46,47, 50, 59,60,63,72,73,76$, $85,86,89$, which are self-orthogonal linear $[42,6,18]$-codes over $\mathbb{F}_3$ with the following Hamming weight enumerator:
$$W_{\mathcal{C}}(Y)=1+14Y^{18}+294Y^{24}+336Y^{30}+84Y^{36}.$$

\section{Conclusion}
Let $G_{(m,3,r)}$ be a metacyclic group of order $3m$, $r\equiv q^\epsilon$ (mod $m$) for some positive integer $\epsilon$ and ${\rm gcd}(m,q)=1$.
We present a system theory of left $G_{(m,3,r)}$-codes over $\mathbb{F}_q$, only using finite field theory and basic theory of cyclic codes and skew cyclic codes. We prove that any left $G_{(m,3,r)}$-code
is a direct sum of concatenated codes with inner codes ${\cal A}_i$ and outer codes $C_i$, where ${\cal A}_i$ is a minimal cyclic code over $\mathbb{F}_q$ of length $m$ and $C_i$ is a skew cyclic code  of length $3$ over an extension field of $\mathbb{F}_q$, and provide
an explicit expression for each outer code in every concatenated code. Moreover, we give the dual code of each
left $G_{(m,3,r)}$-code and determine all self-orthogonal left $G_{(m,3,r)}$-codes over $\mathbb{F}_q$.

% or


\section*{Acknowledgment}
Part of this work was done when Yonglin Cao was visiting Chern Institute of Mathematics, Nankai University, Tianjin, China. Yonglin Cao would like to thank the institution for the kind hospitality. This research is supported in part by the National Natural Science Foundation of China (Grant Nos. 11471255, 61171082, 61571243) and the National Key Basic Research Program of China (Grant No. 2013CB834204).
\end{document}